# Preludes to dark energy:

# Zero-point energy and vacuum speculations

Helge Kragh[*]

**Abstract**  According to modern physics and cosmology, the universe expands at an increasing rate as the result of a "dark energy" that characterizes empty space. Although dark energy is a modern concept, some elements in it can be traced back to the early part of the twentieth century. This paper examines the origin of the idea of zero-point energy and in particular how it appeared in a cosmological context in a hypothesis proposed by Walther Nernst in 1916. The hypothesis of a zero-point vacuum energy attracted some attention in the 1920s, but without attempts to relate it to the cosmological constant that was discussed by Georges Lemaître in particular. Only in the late 1960s was it recognized that there is a connection between the cosmological constant and the quantum vacuum. As seen in retrospect, many of the steps that eventually led to the insight of a kind of dark energy occurred isolated and uncoordinated.

*Content*

[*] Centre for Science Studies, Aarhus University, 8000 Aarhus, Denmark. E-mail: helge.kragh@ivs.au.dk. Manuscript submitted to *Archive for History of Exact Sciences*.

## 1 Introduction

Since the late 1990s, when observations of type Ia supernovae showed the expansion of the universe to be increasing, the "dark energy" supposed to be responsible for the acceleration has been a hot topic in physics and cosmology. The standard view is that the dark energy – a name coined in connection with the discovery[1] – is a manifestation of the cosmological constant introduced by Einstein nearly a century ago. There is another side of the new of mysterious form of cosmic energy. From the perspective of quantum field theory, empty space is characterized by a "zero-point energy" which has the property that its associated pressure is negative and thus makes space expand. To many physicists and cosmologists, the cosmological constant and the zero-point energy density of vacuum are just two names for the same thing.[2]

This paper is not really about dark energy, but it deals with what might be called the prehistory of some of the key elements that eventually coalesced into the modern concept of dark vacuum energy. In this sense, namely that an account of the past is structured and selected with an eye on the present, it may be said to be teleological or "presentist" history of science. On the other hand, I do not allow present knowledge to interfere with my description of the events of the past and thus do not violate accepted historiographical standards.

The chronology of the paper is largely limited to the period from about 1910 to 1935, a period which can with some justification be called the childhood of vacuum energy. It is possible to trace the concept further back in time, even to the days of Newton.[3] However, if one wants to point to pre-quantum and pre-relativity analogies to dark energy, a more sensible arena might be the ethereal world view of the late nineteenth century. The general idea that cosmic space is permeated by an unusual form of hidden energy – a dark energy of some sort – was popular during the Victorian era, where space was often identified with the ether. The generally accepted ethereal medium existed in many forms, some of them assuming the ether to be imponderable while others assumed that it was quasi-material and only differed in degree from ordinary matter. The ether was sometimes thought of as a very tenuous, primordial gas. According to the vortex theory, cultivated by British physicists in particular, the discreteness of matter (atoms) was epiphenomenal, derived from stable dynamic configurations of a perfect fluid. This all-pervading fluid was usually identified with the continuous ether. The highly ambitious vortex theory was not only a theory of atoms, it was a universal theory of ether (or space) and matter, indeed of everything.[4]

---

[1] The term "dark energy" may first have appeared in the title of a scientific paper in Huterer and Turner (1999).

[2] Peebles and Ratra (2003) is a comprehensive and historically informative review. See also the critical analysis in Rugh and Zinkernagel (2002), where the concept of vacuum energy is dealt with from the perspective of philosophy of science.

The point is that by the turn of the nineteenth century few physicists thought of "empty space" as really empty, but rather as filled with an active ethereal medium. H. A. Lorentz and other physicists in the early twentieth century often spoke of the ether as equivalent to a vacuum, but it was a vacuum that was far from nothingness.[5] Although Lorentz was careful to separate ether and matter, his ether was "the seat of an electromagnetic field with its energy and its vibrations, … [and] endowed with a certain degree of substantiality."[6] On the other hand, the popular belief in a dynamically active ether was rarely considered in astronomical or cosmological contexts.

---

[3] Calder and Lahav (2008).
[4] For a full account of the vortex theory, see Kragh (2002).
[5] Examples are given in Illy (1981).
[6] Lorentz (1909, p. 230).

Among the firm believers in the ether as a storehouse of potential energy was the English physicist Oliver Lodge, who has been called a "remote ancestor" of the modern quantum vacuum (see also Section 4).[7] As another example, perhaps an even more dubious ancestor, consider the French psychologist and amateur physicist Gustave LeBon, the discoverer of "black light" and author of the best-selling *The Evolution of Matter*. In this time-typical and hugely popular book, LeBon pictured electrons and other charged particles as intermediates between ordinary matter and the ether. They were "the last stage but one of the disappearance of matter," the last stage being represented by "the vibrations of the ether." Matter formed by electric particles would eventually radiate away all their stored energy and return to "the primitive ether whence they came … [and which] represents the final nirvana to which all things return after a more or less ephemeral existence."[8]

Analogies and precursors apart, in this paper I start (sections 2 and 3) with examining the concept of zero-point energy as it first appeared in Max Planck's so-called second quantum theory of 1911. Although Planck's theory failed to win general approval, the associated hypothesis of zero-point energy of atomic oscillators remained alive. From about 1920 the hypothesis received unexpected support from the half-integral quantum numbers that turned up experimentally in spectroscopy and were eventually justified by the new quantum mechanics. However, in this early period the zero-point energy, if real, was considered a property of material systems and not of empty space. In Section 4 I turn to a remarkable exception from this state of affairs, Walther Nernst's unorthodox, yet in some ways prophetic, theory of a cosmic ether filled with a huge amount of zero-point energy. Nernst's theory is sometimes mentioned by modern physicists, but rarely taken seriously or placed in its proper historical context.[9] Although Nernst's ideas did not make much of an impact on mainstream physics, they inspired a few German physicists to apply quantum theory and thermodynamic reasoning to the universe at large. Works by Otto Stern and Wilhelm Lenz are particularly interesting, and these are dealt with in Section 5.

---

[7] Rowlands (1990, p. 285), a biographer of Lodge, comments: "The infinite energy density of the zero-point vacuum field fluctuations is almost indistinguishable from the infinite elasticity of the universal ethereal medium." Indeed, modern physicists sometimes speak of gravity as the "elasticity of the vacuum" and relate the quantum vacuum to the ether. According to Paul Davies (1982, p. 582), late-nineteenth century physicists "would surely have been gratified to learn that in its modern quantum form, the ether has materialised at last." For other suggestions that the ether has been resuscitated in modern theories of the vacuum, see Wilczek (1999) and Barone (2004).

[8] LeBon (1905, p. 313 and p. 315).

Finally, in Section 6 I turn to the cosmological scene of the 1920s and 1930s. Following a brief consideration of Einstein's resurrection of the ether, I look at how a few physicists came to realize that the cosmological constant can be understood as a vacuum energy density. That this is the case is not surprising from a formal point of view, yet the insight was only spelled out in an address Georges Lemaître gave in 1933. As to the later development, leading to a connection between the cosmological constant and the vacuum of quantum field theory, I only sketch a few of its steps.

## 2  Planck's second quantum theory

The concept of zero-point energy has its roots in a reformulation of the original version of quantum theory proposed by Max Planck in 1900. The revised version was presented in a series of works from 1911 to 1913. Planck first introduced his new radiation hypothesis or "second theory" in an address to the German Physical Society of 3 February 1911, and he subsequently developed it in several papers and lectures, including his report on heat radiation delivered to the first Solvay conference taking place in Brussels from 30 October to 3 November 1911. The new theory became more widely known from the exposition which appeared in the second edition of Planck's *Theorie der Wärmestrahlung* published in early 1913.[10]

---

[9]  Sciama (1978) was among the first to call attention to Nernst's work and its similarity to modern views of the vacuum. See also Sciama (1991, p. 140), where he mentions Nernst's theory, but only "parenthetically" and without providing it with a reference.

Whereas Planck in his original theory of 1900 had treated emission and absorption of radiation symmetrically, in his second theory – at the time generally known as the "theory of quantum emission" – he assumed that only the emission of radiation occurred in discrete energy quanta. The electrodynamic emission of these quanta would be governed by a probabilistic law. Absorption, on the other hand, was supposed to occur in accordance with classical theory, that is, continuously. This feature appealed to physicists who considered Planck's original theory a too radical break with classical physics. For example, in an address of December 1912 Robert Millikan judged the new theory to be "the most fundamental and the least revolutionary form of quantum theory, since it modifies classical theory only in the assumption of discontinuities in *time*, but not in *space*, in the emission (not in the absorption) of radiant energy."[11] Although the radiation was emitted with discrete energy values in Planck's theory, and all of the energy emitted at once, the oscillators did not possess intrinsically discontinuous energies. They could take on any energy, but the emission would only occur when the energy had reached values of $nh\nu$, where $n$ is an integer. As Planck admitted in a letter to Paul Ehrenfest of 23 May 1913: "I fear that your hatred of the zero-point energy extends to the electrodynamic emission hypothesis that I introduced and that leads to it. But what's to be done? For my part, *I hate discontinuity of energy*

---

[10]  Planck (1911). Planck (1912a) and (1912b). Planck (1913), pp. 132-145. Most of Planck's articles on radiation and quantum theory are conveniently collected in Planck (1958). For historical studies of Planck's second theory, see Kuhn (1978, pp. 235-254), Needell (1980), and Mehra and Rechenberg (1982-2000, vol. 1, pp. 124-127, 146-150). See also Darrigol (1988, pp. 63-66).

[11]  Millikan (1913, p. 123).

*even more than discontinuity of emission.*"[12] As stated in the letter, a new and mysterious "zero-point energy" was part and parcel of Planck's new theory.

Based on the ideas underlying the second theory, Planck calculated the average energy of an oscillator vibrating with frequency ν to vary with the absolute temperature $T$ as

$$\bar{E} = \frac{h\nu}{2} \frac{\exp\left(\frac{h\nu}{kT}\right) + 1}{\exp\left(\frac{h\nu}{kT}\right) - 1} ,$$

or

$$\bar{E} = \frac{h\nu}{2} + \frac{h\nu}{\exp\left(\frac{h\nu}{kT}\right) - 1}$$

The values of the quantized energy levels of an oscillator can thus be written

$$\bar{E} = \frac{1}{2}(E_n - E_{n-1}) = \left(n + \frac{1}{2}\right)h\nu$$

where $n$ = 0, 1, 2, … As Planck pointed out, this result implies that at $T$ = 0 (or for $T \rightarrow 0$) the average energy is not zero but equals the finite energy ½$h$ν: "This rest-energy remains with the oscillator, on the average, at the absolute zero of temperature. It [the oscillator] cannot lose it, for it does not emit energy so long that $\bar{U}$ [= $\bar{E}$] is smaller than $h$ν."[13] In order to derive the experimentally confirmed radiation law relating the energy density $\rho$ to frequency and temperature, Planck appealed to the classical limit given by the Rayleigh-Jeans expression

---

[12] Quoted in Kuhn (1978, p. 253). Emphasis added.
[13] Planck (1911, p. 145).

$$\rho(\nu, T) = 8\pi \left(\frac{\nu}{c}\right)^3 \frac{kT}{\nu} \ ,$$

where $k$ is Boltzmann's constant. By making use of a correspondence argument, he obtained in this way the same expression he had derived in 1900:

$$\rho(\nu, T) = \frac{8\pi h\nu^3/c^3}{\exp\left(\frac{h\nu}{kT}\right) - 1}$$

According to Max Jammer, Planck's reasoning in 1911 "was probably the earliest instance in quantum theory of applying what more than ten years later became known as the 'correspondence principle'."[14]

In his *Theorie der Wärmestrahlung*, Planck emphasized that the existence of a zero-point energy was completely foreign to classical physics. However, it seemed to be a ghost-like entity which it was difficult to connect to experiments. As he noted in his first paper of 1911, since the new energy expression of an oscillator differed from the old one by only an additive constant, it would have no effect on the spectrum or on the specific heat as given by $c = \partial \bar{E}/\partial T$. For this reason, Walther Nernst's recent confirmation of Einstein's 1907 theory of the specific heat of solids could not be used to differentiate between the two radiation hypotheses. "Thus, so far it appears not really possible to make a direct experimental test of the new expression for $\bar{U}$ [$= \bar{E}$]," he commented.[15]

---

[14] Jammer (1966, p. 50). Bohr formulated his correspondence principle in 1918, but only used the name (*Korrespondenzprinzip*) in 1920. On the relation between Planck's second theory and Bohr's correspondence principle, see Whitaker (1985) who argues that Planck was the first to make "active use" of correspondence arguments in quantum theory. See also Kuhn (1978, p. 240).

[15] Planck (1911, p. 146).

Planck similarly pointed out that Einstein's controversial theory of light quanta, or rather the photoelectric law derived from it, was unable to distinguish between the two hypotheses.[16] Although he did not think of the zero-point energy as a measurable quantity, or one which would otherwise have direct experimental consequences, he did mention various phenomena that in a qualitative sense might justify it empirically. Among these phenomena was the experimental fact that the energy released in radioactive decay remained uninfluenced by even the most extreme cold. Moreover, the relativistic mass-energy equivalence $E = mc^2$ led naturally to the assumption of "a very considerable intra-atomic amount of energy also a zero absolute temperature."[17]

Planck's second quantum theory was short-lived, a major reason for its short life being its failure to comply with Bohr's atomic theory of 1913 in which both emission and absorption of radiation occurred discontinuously. The successful use of Bohr's theory to atomic and molecular spectroscopy spoke against Planck's second theory which nonetheless may have inspired Bohr in the development of his ideas of atomic structure.[18] At any rate, Bohr soon came to the conclusion that Planck's notion of atomic oscillators was foreign to his atomic theory. In the conclusion of the third part of the trilogy he expressed his misgivings about the Planckian oscillators because they were "inconsistent with Rutherford's theory, according to which all the forces between the particles of an atomic system vary inversely as the square of the distance apart."[19] After all, as he said in a lecture in Copenhagen at the end of 1913, "No one has ever seen a Planck's resonator, nor indeed even measured its frequency of oscillation; we can observe only the period of oscillation of the radiation which is permitted."[20]

---

[16]  Planck (1958, p. 284). See also Wheaton (1983, pp. 178-180).

[17]  Planck (1913, p. 140).

[18]  On the relationship between Planck's second theory of quanta and Bohr's atomic theory, see Hirosige and Nisio (1964), according to whom Planck's revised theory was of great importance to Bohr's original formulation of his theory, as stated in the first part of the 1913 trilogy. On this matter Heilbron and Kuhn (1969, pp. 268-269) disagreed, suggesting that Planck's papers were of no special importance to Bohr. However, Kuhn (1978, p. 320) later admitted that the emission mechanism of Planck's second theory was most likely of relevance to Bohr's thinking.

In an important but unpublished paper of 1916 Bohr emphasized that Planck's second theory was inconsistent with the basic assumption that an atomic system can exist only in a series of discrete stationary states. He argued that the probability of a quantum system being in a state $n$ was given by

$$P_n = n^{r-1},$$

where $r$ denotes the number of degrees of freedom. For a system of several degrees of freedom ($r > 1$), the probability of the system being in state $n = 0$ should thus be zero. "Such a consideration gives a simple explanation of the mysterious zero-point energy," he wrote to the Swedish physicist Carl Wilhelm Oseen.[21] In his unpublished paper he regained the result that at $T = 0$ a harmonic oscillator of two degrees of freedom would have a non-zero energy, but "This so-called zero-point energy has here an origin quite distinct from that in Planck's theory." Bohr elaborated: "In the present theory it arises simply from the fact, that … there is no probability of a periodic system of several degrees of freedom being in the state corresponding to $n = 0$. … At $T = 0$ all the systems are therefore in a state corresponding to $n = 1$."[22]

---

[19]  Bohr (1913, p. 874). In the first part of the trilogy Bohr referred to Planck (1911) and Planck (1912a). He was also acquainted with the Solvay proceedings and thus with Planck (1912b).

[20]  Bohr (1922, p. 10), a translation of an address given to the Danish Physical Society on 20 December 1913 and published in Danish in *Fysisk Tidsskrift* 12 (1914): 97-114.

[21]  Bohr to Oseen, 20 December 1915, in Bohr (1981, p. 567).

Bohr not only applied his theory to the specific heat of hydrogen at low temperatures, but also to the quantized hydrogen atom he had introduced in his theory of 1913. According to this theory, the energy levels of the hydrogen atom were given by

$$E_n = -\frac{2\pi^2 e^4 m}{h^2} \frac{1}{n^2} ,$$

where $e$ is the charge of the electron and $m$ its mass. Since the system has three degrees of freedom, $P_n = n^2$. Bohr explained: "This system affords a peculiar case of zero-point energy. Strictly there is no sense in considering the state corresponding to $n = 0$, since this would correspond to an infinite negative value for the energy; and in order to obtain agreement with experiments it must be assumed that the normal state of the system corresponds to $n = 1$."[23]

At about 1920 few physicists considered Planck's second theory a viable alternative. In a contribution to a special issue of *Die Naturwissenschaften* celebrating the ten-year anniversary of Bohr's atomic theory, Planck admitted that "This second formulation of the quantum theory may be considered today, at least in its extreme form, as finally disproved."[24] What persuaded him was the Stern-Gerlach experiment, which he and most other physicists saw as proof of the discrete stationary states postulated by Bohr's theory. Yet, although the second theory had been abandoned by 1923, one element associated with it continued to live on: the zero-point energy. Planck had himself replaced his second quantum theory with a modified "third theory," and in this version the zero-point energy survived. In a letter of 1915 to Heike Kamerlingh Onnes in Leiden he wrote: "I have almost completed an improved formulation of the quantum hypothesis applied to thermal radiation. I am more convinced than ever that zero-point energy is an indispensable element. Indeed, I believe I have the strongest evidence for it."[25]

---

[22] Bohr (1981, p. 456). See also Gearhart (2010, pp. 146-147). The unpublished paper, intended to appear in the April 1916 issue of *Philosophical Magazine*, was entitled "On the Application of the Quantum Theory to Periodic Systems." Due to Arnold Sommerfeld's new formulation of the quantum theory of atoms, Bohr decided to withdraw it shortly before it was to be published. Incidentally, Sommerfeld ignored the zero-point energy, which is not mentioned in any of the editions (1919-1924) of his influential book *Atombau und Spektrallinien*.
[23] Bohr (1981, pp. 459-460).
[24] Planck (1923, p. 537).

## 3 Half-quanta and zero-point energy

The assumption of a zero-point energy attracted much attention in the physics community, although for more than a decade it remained uncertain whether the quantity was physically real or not.[26] Einstein was perhaps the first to come up with a physical argument for its existence, which he did in a paper of early 1913 co-authored by Otto Stern, a young physical chemist who had recently obtained his doctorate in Breslau under Otto Sackur and subsequently joined Einstein as his assistant, first in Prague and then in Zurich. The two authors considered the rotational energy of a diatomic molecule, as given by

$$E_{rot} = \frac{1}{2} J (2\pi\nu)^2 \,,$$

where $J$ is the moment of inertia and $\nu$ the frequency of rotation. For a collection of molecules at fixed temperature they assumed that all molecules would rotate with

---

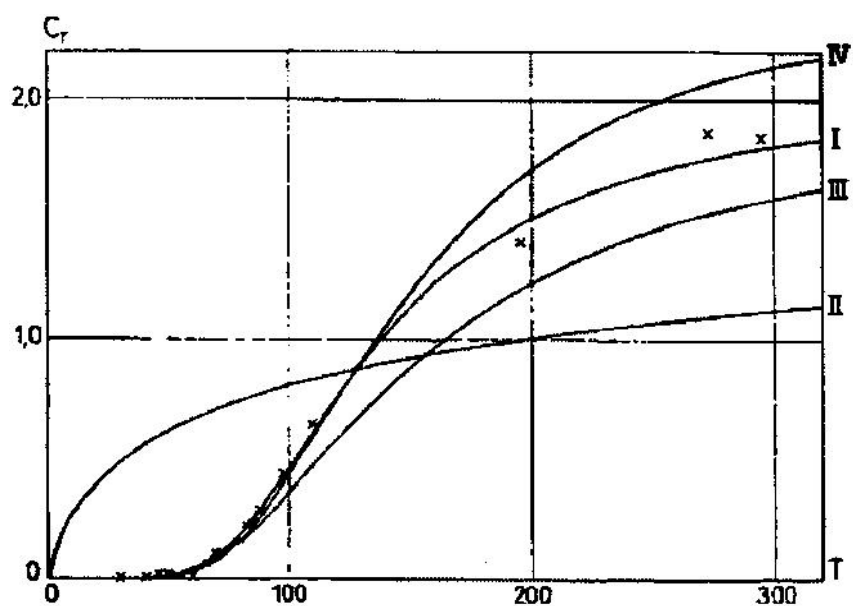

Figure 1. The variation of the specific heat of hydrogen with the absolute temperature, as shown in Einstein and Stern (1913). The crosses are the experimental data obtained by Eucken in the interval from about 30 K to 280 K. The theoretical curve II assumes no zero-point energy, while curve I assumes a zero-point energy of ½hν. The two other curves also assume a zero-point energy, but in curve IV it is equal to hν and in curve III the frequency is assumed to be independent of the temperature.

the same speed. Moreover, they took the rotational kinetic energy to be twice as great as the kinetic energy of a one-dimensional oscillator vibrating at the same frequency ν, or equal to its average energy. From these assumptions they obtained expressions for $c = \partial E_{rot}/\partial T$ in the case of both the first and the second of Planck's hypotheses. Einstein and Stern wanted to establish the different experimental consequences of the two assumptions. This they did by comparing the corresponding specific heats of rotating gas molecules with those measured experimentally. Working at Nernst's laboratory in Berlin, Arnold Eucken had recently obtained data for molecular hydrogen at low temperatures that defied explanation in terms of existing theory.[27] According to the calculations of Einstein and Stern, a fair agreement with Eucken's curve could be obtained if the zero-point energy ½$h\nu$ were included, while Planck's first theory led to quite wrong results (Figure 1). From this followed their cautious conclusion: "Eucken's results on the specific heat of hydrogen make probable the existence of a zero-point energy equal to $h\nu/2$."[28]

---

[27] Eucken (1912).

In spite of the appealing agreement between theory and experiment provided by the assumption of a zero-point energy, Einstein soon retracted his support of it. For one thing, Planck's second quantum theory presupposed harmonic oscillators at fixed frequencies, and there was no reason to expect that it would be applicable also to molecules rotating at frequencies depending on the temperature. Even more problematic was it that Einstein and Stern, by making use of a zero-point energy, were able to derive Planck's radiation law "without recourse to any kind of discontinuities." The problem was not the derivation, of course, but that it relied on a zero-point contribution to the oscillator energy of $h\nu$ and not ½$h\nu$. Moreover, using the value $h\nu$ for rotating molecules spoiled the agreement with Eucken's measurements on the specific heat of hydrogen (Figure 1). The confusion only increased when Paul Ehrenfest in a paper of 1913 showed that he could reproduce the data for low temperatures on the basis of statistical mechanics and Planck's first theory and thus without any zero-point energy.[29]

According to Ehrenfest, the quantum discontinuity was indispensable, whereas the zero-point energy was not. It was in this context that he assumed the angular momentum of the rotator to be quantized according to

$$\frac{1}{2} J(2\pi\nu)^2 = n\frac{h\nu}{2}$$

---

[28] Einstein and Stern (1913, p. 560). In addition to Milloni and Shieh (1991) and Mehra and Rechenberg (1999), see also Gearhart (2010) and Einstein (1995, pp. 270-273).

[29] Ehrenfest 1913. For historical analysis, see Klein (1970, pp. 264-273), Navarro and Pérez (2006, pp. 215-223), and Gearhart (2010, pp. 135-138). In his unpublished paper of 1916, Bohr criticized Ehrenfest's theory and derived from his own theory a curve for the $c$ ($T$) variation that agreed better with the data than the one obtained by Ehrenfest. See Bohr (1981, pp. 458-460).

However, Ehrenfest did not formulate the quantization of angular momentum as a general principle, such as Bohr would do independently a few months later (and as John Nicholson had done in 1912, apparently without Ehrenfest being aware of it).

Already in the fall of 1913 Einstein withdrew his support of the zero-point energy and the results reported in his paper with Stern. During the second Solvay conference in late October 1913 the question of the zero-point energy was discussed by Einstein, Wien, Nernst, and Lorentz. Einstein commented: "I no longer consider the arguments for the existence of zero-point energy that I and Mr. Stern put forward to be correct. Further pursuit of the arguments that we used in the derivation of Planck's radiation law showed that this road, based on the hypothesis of zero-point energy, leads to contradictions."[30] In a letter to Ehrenfest a few days later he declared the zero-point energy "dead as a doornail" (*Mausetot*).[31] However, the announcement of death was premature.

Two points with regard to the paper by Einstein and Stern should be emphasized. First, they did not quantize the rotator, but allowed it to have a continuum of energies depending on the temperature. Second, they only attributed a zero-point energy to material objects, either oscillating electrons or rotating diatomic molecules, while they did not apply the additional energy term to the electromagnetic field. In retrospect this explains how they were able to derive the correct Planck spectrum on the basis of the wrong zero-point energy $h\nu$. This value happens to be the correct one for the *sum* of the interacting harmonic oscillators and the energy of the electromagnetic field.[32]

---

[30] Einstein (1995, p. 553).
[31] Einstein to Ehrenfest, before 7 November 1913, and also Einstein to Ludwig Hopf, 2 November 1913: "One hopes Debije [Debye] will soon demonstrate the incorrectness of the hypothesis of zero-point energy, the theoretical untenability of which became glaringly obvious to me soon after the publication of the paper I coauthored with Mr. Stern." Both letters are reproduced in Einstein (1993, pp. 563-565).

Far from being dead as a doornail, after 1913 the zero-point energy continued to attract a great deal of attention among physicists and physical chemists, in many cases independent of Planck's second theory. A possible way to answer the question of a zero-point motion might be to study the X-ray diffraction pattern in crystals at low temperature. If the atoms in a crystal had a zero-point motion, this would presumably influence the intensity distribution in the diffraction pattern. This line of research, which eventually led to a "direct proof" of zero-point motion, was pioneered by Peter Debye in studies of 1913-1914, but without leading to a conclusive answer.[33]

Another line of research was related to the attempts to separate isotopes by chemical means or fractional distillation. The British physicist Fredrick Lindemann, a former collaborator of Nernst, showed that in principle such separation would be possible, but that it would depend on whether or not there was a zero-point energy: "The amount of separation to be expected depends upon … whether 'Nullpunktsenergie' is assumed. … The difference should be measurable if there is no 'Nullpunktsenergie', and it is suggested that experiments on the vapour pressure and affinity of isotopes would give valuable information on this important point."[34] If a zero-point energy were not assumed, the expected separation effect would be tiny. However, experiments of the kind proposed were unable to settle the question and tell whether the zero-point energy existed or not.

---

[32]  See Milloni and Shih (1991), who discuss the question and what reasons Einstein and Stern might have had to ignore the zero-point energy of the field. The shortcomings of the Einstein-Stern paper are also discussed in Sciama 1991.

[33]  Debye (1913). Debye (1914). See also Mehra and Rechenberg (1982-2000, vol. 5, pp. 143-146).

[34]  Lindemann (1919, p. 181).

Arguments somewhat similar to Lindemann's were a few years later suggested by Stern, who discussed them with a skeptical Pauli. In a letter of 1960, Stern recalled:

> Pauli and I continually discussed the question of the zero-point energy in Hamburg in the early 1920s. … I for my part always tried to convert Pauli to the zero-point energy against which he had the gravest hesitations. My main argument was that I had calculated the vapour pressure differences of the neon isotopes 20 and 22, which Aston had tried in vain to separate by distillation. If one calculates without zero-point energy there results such a large difference that the separation should have been quite easy. The argument seemed (and seems) to me so strong because one does not assume anything else than Planck's formula and the fact that isotopes are distinguished only by the atomic weight.[35]

However, at the time Pauli remained unconvinced. As mentioned by Stern, as early as 1913 he had studied the vapor pressure of monatomic gases and arrived at an expression for the heat of vaporization which he interpreted as support of a zero-point energy.[36] He gave a more elaborate version in a work of 1919, in which he calculated the vapor pressure above the surface of a solid body, which he conceived as a collection of $N$ atoms vibrating harmonically in three dimensions with frequencies $\nu_k$. To obtain agreement with experimental data he suggested that the heat of vaporization at $T = 0$ was smaller than the potential energy of the $N$ atoms in the gaseous state. That is, in the solid equilibrium state the atoms were not at rest, but possessed a vibrational energy of

---

[35] Stern to Enz, 21 January 1960, in Enz (2002, p. 150). See also Enz (1974), reprinted in Enz (2009, pp. 63-72). Stern did not publish his calculations on the isotopic effect.
[36] Stern (1913).

$$E_0 = \frac{1}{2} h \sum_1^{3N} \nu_k$$

Stern expressed his hope that the zero-point energy of solid bodies would find its interpretation in "The more recent works of N. Bohr [in which] this hypothesis in a somewhat modified form has acquired a very deep meaning."[37]

Rather than considering vaporization, the two young physical chemists Kurt Bennewitz and Franz Simon (later Sir Francis Simon), who worked at Nernst's laboratory in Berlin, studied the melting process at low temperatures. Their complex calculations of the melting points of hydrogen, argon and mercury led them to conclude that the results provided evidence for a zero-point energy. Moreover, they suggested – correctly, as it later turned out – that this quantity was responsible for the difficulty in solidifying helium even at very low temperature.[38] According to Bennewitz and Simon, the zero-point energy in liquid helium would act as an internal pressure, expanding it to such a low density that no rigid structure of the atoms could be maintained.

Among the early and most persistent advocates of the zero-point energy was the Dutch physicist Willem Keesom, who at the 1913 Wolfkehl meeting in Göttingen defended the new Einstein-Stern theory and suggested that the zero-point energy might also turn up in the equation of state of monatomic gases.[39] Several other speakers at the Wolfkehl meeting commented on the zero-point energy, including Planck, Kamerlingh Onnes, Debye, and Sommerfeld. For a while the subject was taken seriously among Dutch physicists in Leiden and Utrecht, where it came up in particular in connection with research in magnetism at low temperature.[40] As Keesom saw it, the evidence in favor of zero-point energy was far stronger than the counterevidence. Yet, evidence is not proof, and in the decade after 1911 the problem remained unresolved. As mentioned, the abandonment of Planck's second theory did not imply that the idea of zero-point energy was abandoned.

---

[37]  Stern (1919, p. 77). It is unclear which works of Bohr Stern had in mind, but he may have thought of the correspondence principle. In fact, Bohr did not deal with zero-point energy in any of his published papers 1913-1919.

[38]  Bennewitz and Simon (1923). The section on zero-point energy was written by Simon.

[39]  Planck et al. (1914, p. 166 and p. 194). Keesom (1913).

Einstein would have nothing of it. "It is well known that all theories characterized by a 'zero-point energy' face great difficulties when it comes to an exact treatment," he wrote in a paper of 1915. "No theoretician," he continued, "can at present utter the word 'zero-point energy' without breaking into a half-embarrassed, half-ironic smile."[41] Yet several years later Einstein returned to the question, now with a more sympathetic view. In his correspondence with Ehrenfest from 1921-1923 he suggested that the zero-point energy might play a role in the cases of hydrogen and helium. Perhaps, he suggested, it might explain the density maximum in helium. However, neither Einstein nor Ehrenfest turned their ideas on the subject into publications.[42]

According to Bohr's atomic theory quantum numbers had to be integers, but by the early 1920s a growing amount of evidence indicated that in some cases "half-quanta" of the kind first considered by Planck in 1911 had to be accepted. These half-integral quantum numbers first turned up in attempts to understand the band spectra emitted by molecules. In 1919 Elmer Imes at the University of Michigan published precision experiments on the absorption of HCl and HBr that showed a distinct gap in the centre of the pattern of lines.[43] In order to explain Imes's data, the Berlin physicist Fritz Reiche suggested changing the standard rule for rotational quantization, namely by changing the formula for the energy of a rotator from

$$E_{rot} = m^2 \frac{h^2}{8\pi^2 J} \qquad \text{to} \qquad E_{rot} = (m+\frac{1}{2})^2 \frac{h^2}{8\pi^2 J} \; ,$$

which, since $m = 0, 1, 2, \ldots$, implied a zero-point rotational energy. This conclusion, that a diatomic molecule cannot exist in a rotation-free state, he justified by Bohr's new correspondence principle.[44] According to Reiche, the suggestion of rotational half-quanta was first suggested by Einstein, "with whom I have often had the opportunity to discuss these matters, … [and who mentioned] a possible way to change the rotational quantization so as to annul the contradiction with observations."[45] Although the half-quanta were theoretically controversial they seemed necessary and were adopted by several molecular spectroscopists. For example, they were incorporated into an influential and more elaborate theory of band spectra that Adolf Kratzer, a physicist at the University of Münster, published in 1923.[46]

[40]  For a survey of zero-point energy in Leiden, see van Delft (2007, pp. 484-493) and van Delft (2008).
[41]  Einstein (1915, p. 237).
[42]  Einstein to Ehrenfest, 1 September 1921, in Einstein (2009, p. 265). See also excerpts of the Einstein-Ehrenfest correspondence in Mehra and Rechenberg (1982-2000, vol. 1, pp. 571-572).

    With the new studies of band spectra in the early 1920s the concept of zero-point energy became respectable among molecular physicists. Yet it was only in the fall of 1924 that half-quanta were firmly established in molecular spectroscopy. In a study of the spectrum of boron monoxide (BO), Robert

[43] Imes (1919). For the history of "half quanta," see Gearhart (2010).

[44] The same result, also based on the correspondence principle, was derived in Kramers and Pauli (1923).

[45] Reiche (1920, p. 293). See also Reiche (1921, pp. 155-159), which included sections on Planck's second theory and the zero-point energy, which he clearly was in favor of (pp. 32-33).

[46] Kratzer (1923), and see also Barker (1923).

Mulliken, a young physical chemist at Harvard University, concluded that observations could only be understood on the assumption of quantum numbers with a minimum value of one half. In a preliminary announcement of his results in *Nature*, he wrote:

> It is probable that the minimum vibrational energy of BO (and doubtless of other) molecules is ½ quantum. In the case of molecular rotational energy, the necessity of using half quanta is already well established. Analogous relations appear in line spectra; e.g. Heisenberg has successfully used half-integral radial and azimuthal quantum numbers in explaining the structure and Zeeman effect of doublets and triplets.[47]

In the full report that appeared in *Physical Review* in March 1925, Mulliken similarly concluded that his work "would involve a null-point energy of ½ quantum each of vibration and rotation" and he related it to Lindemann's investigation of the vapor pressures of isotopes.[48] His paper was widely considered a final confirmation of half-quanta and, by implication, a form of zero-point energy. On the other hand, in spite of being anomalous the result had almost no effect at all on the crisis in quantum theory that a few months later would lead to Heisenberg's formulation of a new quantum mechanics – and thereby to a theoretical justification of the zero-point energy of an oscillator.

   At this place a brief terminological note may be appropriate. What Planck had originally called *Restenergie* (rest energy) soon became known as *Nullpunktsenergie*, a name used by, for example, Einstein and Stern in their paper of 1913. For a while the German term – or sometimes the equivalent "null-point energy" as used by Mulliken – was used also in the English scientific literature, such as exemplified by the papers by Lindemann (1919) and Tolman (1920). Only from about 1925 did it become common to refer to "zero-point energy." This term may first have been used by Bohr in his unpublished paper of 1916.

---

[47]  Mulliken (1924). See also Mehra and Rechenberg (1999).
[48]  Mulliken (1925, p. 281).

As indicated in the quotation from Mulliken, half-quanta also played a role in some of the attempts to understand what was probably the most serious problem in the old quantum theory, namely, the anomalous Zeeman effect. Thus, according to young Heisenberg's so-called core model of the atom, electrons could be in a state given by the azimuthal quantum number $k = \frac{1}{2}$, which was difficult to harmonize with the established Bohr-Sommerfeld atomic model.[49] When Heisenberg introduced half-integral quantum numbers, he was originally unaware of the earlier discussion related to Planck's second theory and the possibility of a zero-point energy. It seems to have been Pauli who directed his attention to this discussion and to Stern's paper of 1919, and Heisenberg also had conversations with Kratzer who informed him about the use of half-quanta in the study of band spectra.[50] To make a long story short, in spite of resistance from Bohr and other leading physicists the evidence for half-quanta and zero-point energy could not be ignored: physicists learned to live with them, if not love them.

Only with the emergence of quantum mechanics did the concept of zero-point energy become really respectable and seen as a consequence of a fundamental physical theory. In his famous *Umdeutung* paper from September 1925 Heisenberg applied his new quantum formalism, soon known as the Göttingen quantum mechanics, to the simple harmonic oscillator. The result of Heisenberg's calculations was that the energy of the oscillator was not limited to the values $E_n = nh\nu$, but instead to

$$E_n = (n + \tfrac{1}{2})h\nu$$

---

[49] For Heisenberg's early atomic model based on half quantum numbers, dating from 1922, see Casssidy (1978) and Cassidy (1979).

[50] Mehra and Rechenberg (1982-2000, vol. 2, p. 30). Gearhart (2010, pp. 160-161).

For the anharmonic oscillator he derived a more complicated expression, also involving a zero-point energy. Although Heisenberg's result was the same as Planck's formula from 1911, there was the difference that in the case of quantum mechanics it is valid also for an individual oscillator and not merely as an average. The result was duplicated by Erwin Schrödinger in the second of his communications on wave mechanics from April 1926, where he commented: "Strangely, our quantum levels are *precisely* the same as in Heisenberg's theory!"[51] The formal equivalence between wave mechanics and the Göttingen quantum mechanis was only proved a month later.

     By the summer of 1926 the zero-point energy was no longer controversial, at least not in so far as it concerned material systems. In the later literature on quantum physics it became customary to see the zero-point energy as a straightforward consequence of Heisenberg's uncertainty principle for position and momentum. If a harmonic oscillator were to have zero energy, both its potential and kinetic energy would have to be zero. The case $E_{pot} = \tfrac{1}{2}kx^2 = 0$ would correspond to a precise knowledge of the position of the particle ($x = 0$) and thus imply that the momentum $p$ is completely uncertain. However, then the mean value of $E_{kin} = p^2/2m$ would be infinite. Conversely, for $E_{kin}$ to be zero, $E_{pot}$ would have to be infinite. A simple calculation shows that the ground state of a quantum harmonic oscillator is equal to the minimum energy allowed by the uncertainty principle $\Delta x \Delta p \geq h/2\pi$, and that this energy is just $\tfrac{1}{2}h\nu$.

---

[51] Heisenberg (1925). Schrödinger (1926, p. 516). Contrary to Heisenberg, Schrödinger noted the connection to the old question of the validity of Planck's second quantum theory.

## 4  Nernst's cosmic quantum-ether

The first suggestion of applying the concept of zero-point energy to free space, and in this way turning it into a tool of possible relevance for cosmological research, came from an unlikely source. The great physical chemist Walther Nernst had established his reputation by pioneering works in electrochemistry and chemical thermodynamics, culminating in 1906 with the heat theorem also known as the third law of thermodynamics.[52] In its original formulation the theorem was a method of calculating free energies and equilibrium constants from calorimetric data, whereas Nernst resisted the later formulation that all entropy differences ∆$S$ vanishes at $T = 0$.[53] It was primarily for his work in thermodynamics he was awarded the 1920 Nobel Prize in chemistry. The new heat theorem led Nernst from chemistry to quantum physics, a move inspired by Einstein's 1907 theory of the specific heats of solids which Nernst confirmed in a series of low-temperature experiments conducted about 1910.

Nernst's debut in quantum theory took place in early 1911, when he submitted a paper on the theory of specific heats in which he applied quantum theory to diatomic gases such as hydrogen. Although Nernst did not quantize the rotating molecule, he did arrive at a quantum-based phenomenological expression for the variation of the specific heat of a diatomic gas with temperature.[54] Later the same year he reported on his formula and related subjects at the memorable first Solvay conference, a meeting of which he was the chief organizer. In Brussels he listened to Planck's exposition of his second quantum hypothesis and its associated concept of zero-point energy. As Nernst suggested in the subsequent discussion, the zero-point energy would imply that at the absolute zero of temperature a solid body would still have a vapor pressure, a claim that Planck however denied.[55]

---

[52]  The literature on Nernst is extensive. For full biographies, see in particular Barkan (1999) and Bartel and Huebener (2007). The scientific works of Nernst are well discussed in Partington (1953). A useful website, with a bibliography of the works of Nernst and his students, is http://www.nernst.de.

[53]  For the complex history of Nernst's heat theorem, see Simon (1956).

[54]  Nernst (1911), discussed in Gearhart (2010).

At about the same time that Nernst entered quantum theory, he took up an interest in cosmological questions (Figure 2). His first excursion into cosmology and cosmic physics was not motivated by quantum theory, but by the old and still much-discussed question of a universal *Wärmetod* (heat death) caused by the ever increasing amount of entropy (or ever decreasing amount of free energy). This supposed consequence of the second law of thermodynamics, first stated in different versions by Hermann von Helmholtz, William Thomson and Rudolf Clausius in the mid-nineteenth century, was highly controversial for both scientific and non-scientific reasons because it predicted the end of the world, or at least the end of all activity and life in the world.[56] Nernst's Swedish colleague in physical chemistry, Svante Arrhenius, were among those who in the early years of the twentieth century resisted the heat death scenario and suggested cosmic mechanisms which would counter the deadly growth in entropy. Probably inspired by Arrhenius's writings, Nernst did the same. Ever since 1886, when he first became acquainted with Boltzmann's gloomy prediction of an unavoidable cosmic heat death, he denied this alleged consequence of the second law of thermodynamics.[57] For him, as for several of his colleagues in science, it was an intellectual necessity to establish a cosmology that secured eternal evolution in an infinite, self-perpetuating universe.

---

[55] Langevin and de Broglie (1912, p. 129).

[56] The history of the heat death and its associated concept of a beginning of the world is detailed in Kragh (2008), which emphasizes the religious and other non-scientific aspects of the controversy. This work gives further references to the literature.

[57] Nernst attended Boltzmann's inaugural lecture of 1886 in Vienna, in which Boltzmann maintained that the heat death followed from thermodynamics (Nernst 1921, p. 1). Historical and other works that deal with Nernst's cosmological and astrophysical views include Kragh (1995), Bartel and Huebener (2007, pp. 306-326), Browne (1995), and Huber and Jaakkola (1995).

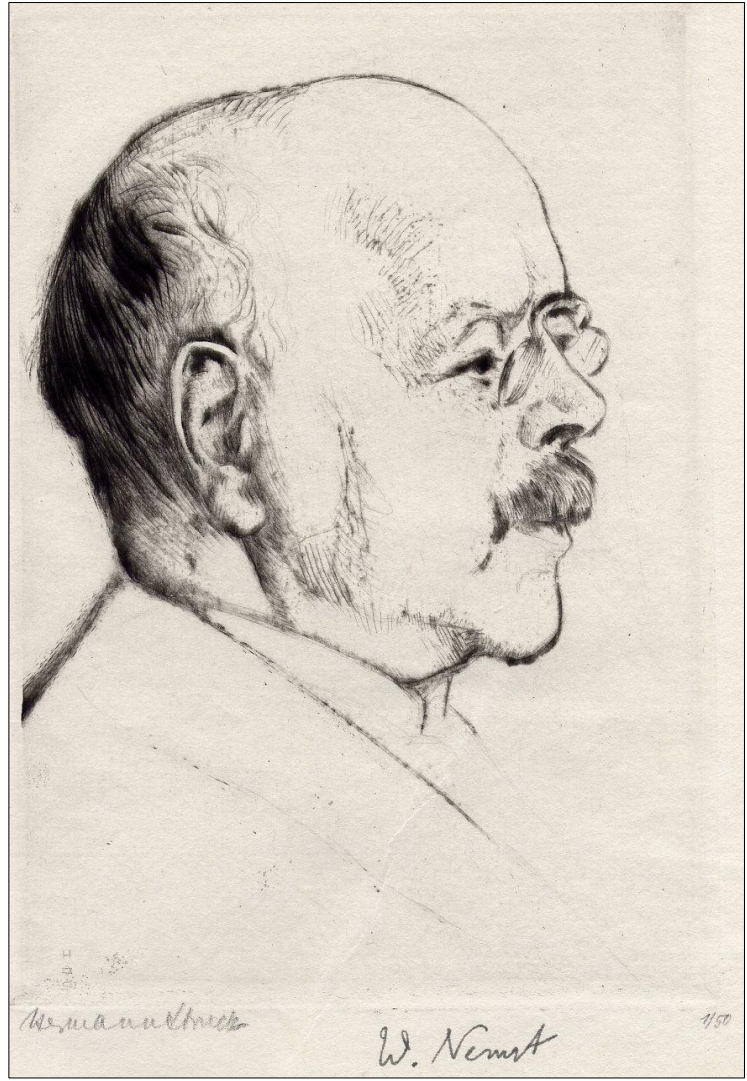

Figure 2.  Walther Nernst at the time he developed his ideas of a universe filled with zero-point radiation. Etching by Hermann Struck from 1921. Source: http://www.nernst.de.

In a lecture given to the 1912 meeting of the Society of German Scientists and Physicians in Münster, Nernst indicated a way in which the world might be saved from the heat death without abandoning the second law of thermodynamics. His tentative solution involved as major ingredients radioactivity and the ether, with the latter supposed to be the ultimate end product of radioactive decay. (As mentioned in Section 1, a similar idea had earlier been entertained by LeBon and other authors.) Like the free energy, radioactivity was known to decrease irreversibly, but alone it would not do in countering the entropy increase. On the contrary, "the theory of radioactive decay of the elements has augmented the above-mentioned degradation of energy with a correspondingly steady degradation of matter, and thus has only doubled the prospects of an Armaggeddon of the universe [*Götterdämmerung des Weltalls*]."[58]

Nernst was not the first one to use radioactivity in a cosmological context. One year earlier the Austrian physicist Arthur Erich Haas had reached the same conclusion, that radioactive decay constituted one more argument for the end of the universe.[59] But whereas this was a conclusion Haas happily welcomed, Nernst thought he was able to circumvent it and turn it into an argument for a static and eternally active world. This is where the ether entered, namely, as a medium that *ex hypothesi* counteracted the degradation of matter and energy. According to Nernst:

> The atoms of all elements of the universe will sooner or later entirely dissolve in some primary substance [*Ursubstanz*], which we would have to identify with the hypothetical medium of the so-called luminiferous ether. In this medium … all possible configurations can presumably occur, even the most improbable ones, and consequently, an atom of some element (most likely one with high atomic weight) would have to be recreated from time to time. … This means, at any rate, that the cessation of all

---

[58]  Nernst (1912, p. 105), with partial translation in Huber and Jaakkola (1995).
[59]  Haas (1912), who shared Nernst's belief that all elements are radioactive. On the role played by radioactivity in cosmology and astrophysics in the early twentieth century, see Kragh (2007).

events no longer needs to follow unconditionally as a consequence of our present view of nature.[60]

Nernst's cosmic hypothesis was admittedly speculative, and he emphasized that it should not be taken as a new cosmological theory but merely as an illustration of what he called "the thermodynamic approach" to the study of the universe. His chief hypothesis, an active ether in constant interaction with matter, was not particularly novel, and in 1912 he did not refer to either quantum theory or zero-point energy. However, in an article four years later he did make the connection.

In this paper, a lengthy communication to the German Physical Society read on 28 January 1916, Nernst proposed that empty space (or ether, as he saw it) was filled with electromagnetic zero-point radiation. Although he found the zero-point energy useful for the energy-rich ether, he was not satisfied with Planck's version of it because he felt it violated the universal validity of electrodynamics. His own alternative, he emphasized, "succeeds in taking over without changes the most important laws of the old physics, [and this] I consider not only an advantage but also a probable reason for admitting it as acceptable."[61] Contrary to Planck and other early researchers, Nernst's zero-point energy was a concept that characterized both material objects (oscillators and rotators) and the radiation filling up the ether: "Even without the existence of radiating matter, that is, matter heated above absolute zero or somehow excited, empty space – or, as we prefer to say, the luminiferous ether – is filled with radiation."[62] The two were interconnected, for a vibrating electron would constantly exchange energy with the zero-point radiation of the ether.

---

[60] Nernst (1912, pp. 105-106).
[61] Nernst (1916, p. 107). For the differences between the zero-point energies of Nernst and Planck, see Enz (1974).
[62] Nernst (1916, p. 86).

Yet another difference between the systems of Planck and Nernst was that whereas the law of energy conservation was strictly valid for the first, according to Nernst it was only statistically valid, "just like the second law of thermodynamics." For a single atom or molecule the energy did not need to be conserved, since the material object would exchange energy with the hidden energy pool of empty space. This was a conception to which Nernst would return a few years later, extending it to the general suggestion that *all* the laws of nature were of a statistical nature. At this occasion, his inaugural lecture as rector for the University of Berlin given on 15 October 1921, he repeated his idea that an enormous amount of energy was stored in the light ether in the form of zero-point energy.[63] Nernst's ether was quasi-material, in the sense that he imagined it to consist of tiny "molecules," neutral doublets made up of two polar particles of "unbelievably small dimensions." This idea of a corpuscular ether was not central to his arguments, however. He merely seems to have reused an older idea of his, namely that the ether consists of weightless combinations of positive and negative electrons. These hypothetical particles he called "neutrons."[64]

It was an important part of Nernst's hypothesis that calculations of the zero-point energy followed from the ordinary theory of statistical mechanics if only the quantity $kT$ were replaced by $h\nu$. This implies that for each degree of freedom, where classical theory assigns the energy $\frac{1}{2}kT$ the zero-point energy

---

[63] Nernst (1922, p. 493), with contextual comments in Forman (1971, pp. 84-87). At about the same time, Charles Darwin, Hendrik Kramers, Bohr, and a few other physicists contemplated the idea that strict energy conservation might break down in the interaction between radiation and matter, a view which explicitly appeared in the Bohr-Kramers-Slater theory of 1924.

[64] Nernst (1916, p. 110). Nernst's neutronic ether appeared in Nernst (1907, p. 392) and also in later editions of his textbook, for example in the 15th edition of 1926 (pp. 464-465), where he stated that the electrons making up the neutrons would become ponderable by taking up zero-point energy. He may have taken the idea, as well as the name "neutron," from Sutherland (1899). Nernst's neutron had only the name in common with the neutron that Rutherford introduced in 1920 as a material proton-electron composite particle.

becomes ½ $h\nu$. For example, the ground state of a one-dimensional oscillator becomes $h\nu$ and not, as in Planck's theory, ½ $h\nu$.[65] He commented: "Every atom, and likewise every conglomerate of atoms, which is capable of oscillation at a frequency $\nu$ per second owing to its mechanical conditions, will per degree of freedom take up the kinetic energy $E$ = ½ $h\nu$ and that even, as already noted, at the absolute zero. … Contrary to the usual heat motion, but in accordance with thermodynamics, the zero-point energy is, like every other form of energy at absolute zero, free energy."[66] In the case of the energy density of the zero-point radiation at frequency $\nu$, Nernst adopted the formula

$$\rho(\nu, T) = \frac{8\pi h}{c^3} \nu^3 \,,$$

which derives from the classical Rayleigh-Jeans law by replacing $kT$ with $h\nu$. The total energy density integrated over all frequencies in the range from zero to infinity then becomes infinite. Although Nernst saw "no reason to call such a conception impossible," of course he realized that an infinite energy density is unphysical. Based on his idea of an atomistic ether, he therefore considered to replace the $\nu^3$ law with the expression

$$\rho(\nu, T) = \frac{8\pi h\nu^3}{c^3 \nu_0} \frac{\nu}{\exp(\nu/\nu_0) - 1} \,,$$

where $\nu_0$ is a constant characteristic of the structure of the ether-vacuum. However, given the lack of knowledge of the value of $\nu_0$ he chose to return to the

---

[65] Nernst (1916, p. 87). See also Peebles and Ratra (2003, p. 571).
[66] Nernst (1916, pp. 86-87).

$\nu^3$ law and provide it with a cut-off corresponding to some maximum frequency $\nu_m$. The result becomes

$$\rho = \int\limits_{0}^{\nu_m} \frac{8\pi h}{c^3} \nu^3 d\nu = \frac{2\pi h}{c^3} \nu_m^4$$

Nernst assumed $\nu_m = 10^{20}$ Hz, or $\lambda_{min} = 3 \times 10^{-10}$ cm, and with this value he obtained a lower limit for the energy density, namely

$$\rho = 1.52 \times 10^{23} \text{ erg/cm}^3$$

In modern units the quantity is equal to $1.52 \times 10^{16}$ J/cm$^3$ $\cong 10^{29}$ MeV/cm$^3$ or, by $E = mc^2$, about 150 g/cm$^3$. "The amount of zero-point energy in the vacuum is thus quite enormous, making extraordinary fluctuations in it to exert great actions," he wrote.[67] Referring to a result obtained by Planck for the energy density of heat radiation, Nernst further showed that if a zero-point radiation enclosed in a container is compressed, neither its energy density nor its spectral distribution will be affected: "Any doubts one might raise to the zero-point radiation owing to radiation pressure or resistance to bodies moving through the vacuum are overcome by this truly remarkable [*gewiss merkwürdige*] result."[68] The remarkable result relied on the relationship $\rho \sim \nu^3$, for which reason Nernst considered it to support his theory. The invariance of the energy density would later reappear as a property of the "false vacuum" of inflation cosmology and, even later, of dark energy.

---

[67]  Nernst (1916, p. 89).
[68]  Nernst (1916, p. 90).

Although it is Nernst's cosmophysical speculations based on an ethereal zero-point energy that are of interest in the present context, these ideas played only a limited role in his 1916 essay. The main part of it was concerned with more mundane applications, in particular to chemical reaction rates, equilibrium processes, and the structure of the hydrogen molecule. Based on his zero-point version of quantum theory he proposed a model of the hydrogen molecule that in some respects differed from the Bohr-Debye model generally accepted at the time.[69] According to this model, the two revolving electrons were placed opposite on a circular orbit perpendicular to and between the two hydrogen nuclei. Nernst ascribed to each of the two electrons a kinetic theory of $\frac{1}{2}h\nu$, where $\nu$ is the frequency of revolution. Because the electrons were in equilibrium with the zero-point radiation, they would not radiate, which explained the stability of the model without sacrificing the validity of ordinary electrodynamics as postulated by Bohr. For the moment of inertia of the hydrogen molecule Nernst derived $J = 3.6 \times 10^{-41}$ g cm$^2$, which he found was in better agreement with measurements than the value used by Debye (which was $1.2 \times 10^{-40}$ g cm$^2$).

In a booklet of 1921, entitled *Das Weltgebäude im Lichte der neueren Forschung* and based on a popular lecture he gave in Berlin, Nernst elaborated on the cosmological and astrophysical consequences of his hypothesis. His larger aim was the same, to demonstrate that eternal matter-ether recycling prevented the heat death and secured a static universe without a beginning or an end: "Our eyes need not, in the far future, have to look at the world as a horrible graveyard, but as a continual abundance of brightly shining stars which come into existence and disappear."[70]

---

[69] Nernst (1916, pp. 104-106). The Bohr-Debye model was essentially Bohr's original model (Bohr 1913) in the improved form which Debye had reported in Debye (1915).

[70] Nernst (1921, p. 37). See also Bromberg (1976, pp. 169-171).

More clearly than earlier Nernst explained that atoms of the chemical elements appeared out of the fluctuations of the ether, and that these atoms or their decay products would again disappear in the zero-point energy of the ethereal sea. This idea also appeared in several of his later works, where he attempted to develop it into a proper theory of astro- and cosmophysics. For example, in his work of 1921 he considered the temperature of cosmic space, as usually identifying empty space with the ether. Without providing a value for the very low temperature, he argued that the ether must have a small capacity for absorbing heat rays and that this absorption of heat would eventually turn up as zero-point energy in the ether. This theme he developed in later works, in 1938 arriving at a cosmic "background temperature" of about 0.75 K, a result he considered to be "not implausible."[71] But we shall not here be concerned with Nernst's cosmological views in the 1930s or with his attempt to interpret Hubble's law of expansion as a quantum effect in a stationary universe.[72]

As Nernst pointed out in his *Weltgebäude* of 1921, the German physicist Emil Wiechert, a pioneer of geophysics and electron theory, had independently arrived at a view of the universe that in many ways was similar to his own. Wiechert adhered to the ether no less fully than Nernst, and his ether was no less physically active and rich in energy. Like Nernst, he speculated that ether-matter transmutations might continually take place in the depths of space, and in this way provide a cosmic cycle that would make the heat death avoidable. According to Wiechert, material atoms were to be seen as extraordinary configurations in the ether, which had to be assigned a content of energy. "With regard to the structure of the electron," he wrote, "it follows that the energy density of the ether must be considered to be comparable to at least $7 \times 10^{30}$ erg/cm$^3$. … One gets an impression of the forces that govern the ether when one recalls that the pressure which comes into play by keeping together the electric charge in an electron is of the order $7 \times 10^{24}$ atmospheres."[73]

---

[71] Nernst (1938). For other early attempts to estimate the temperature of space, or (anachronistically) the temperature of the cosmic background radiation, see Assis and Neves (1995).

[72] For these aspects, see Kragh (1995) and Bartel and Huebener (2007).

Whereas Wiechert did not follow Nernst in making use of the zero-point energy, or otherwise refer to quantum theory, he related the energy of the ether to the cosmological constant appearing in Einstein's field equations of 1917. (Nernst ignored general relativity and never mentioned the cosmological constant or Einstein's world model.) Although strongly opposed to the theory of relativity, in large measure because it disposed of his beloved ether, Wiechert suggested that the general theory had in effect resurrected the ether and that the cosmological constant ($\Lambda$ or $\lambda$) somehow played a role in the resurrection. "My impression is that the $\lambda$-term does not subordinate the ether to matter, but, on the contrary, subordinates matter to ether; for now matter appears as precipitations from the ether which here and there are rolled up and thereby cause insignificant changes in the constitution of the ether."[74] He was not more concrete than that.

Nernst's speculations had some similarity to ideas about the structure and function of the classical ether that for a time survived the relativity and quantum revolutions. As mentioned in Section 2, the physically active ether was a widely accepted component of the Victorian world view, with Oliver Lodge being among the most enthusiastic of its protagonists. Lodge shared some of the cosmological views of Nernst, Wiechert and Millikan, including that matter particles generated from the potential energy of the ether might act counter-entropically and prevent the heat death of the universe. He likewise speculated that radioactivity might not be limited to processes of degeneration, but also involve regeneration of matter.

---

[73] Wiechert (1921a, p. 66). See also Wiechert (1921b).
[74] Wiechert (1921a, p. 69), who was aware of Nernst's ideas. In Wiechert (1921b, p. 186) he referred to Nernst (1916).

Quantum theory had no role to play in Lodge's ether, but otherwise it had a great deal in common with Nernst's. First and foremost, it was filled with an enormous amount of energy that, although not directly detectable, could be calculated. Restating an earlier estimate, in 1920 Lodge concluded that "the ether may quite well contain a linear dimension of the order $10^{-30}$ to $10^{-33}$ centim., and an energy of $10^{30}$ to $10^{33}$ ergs per cubic centimeter."[75] The energy density of Lodge's ether was thus of the same order as the one calculated by Wiechert. It corresponds to about 10,000 tons/cm$^3$.

The ideas that Nernst entertained with regard to ether and zero-point energy seem to have been well known in Germany. However, they did not attract much scientific interest among mainstream physicists, who may not have found his arguments for a vacuum zero-point radiation convincing. The general attitude may rather have been the one summarized by Siegfried Valentiner, professor of physics at the Mining Academy in Clausthal: "It is much more difficult to conceive the presence of such a zero-point energy in the vacuum filled with electrical radiation than it is to assume that the existence of the zero-point energy is a peculiarity of the [material] oscillators."[76]

The cosmological considerations of Nernst were positively reviewed by Paul Günther, a physical chemist and former student of his, and they were disseminated to audiences in both Russia and the United States.[77] Nernst's use of the cosmic zero-point energy as a means to counteract the entropic heat death was occasionally noticed in the philosophical and theological debate concerning the end of the world, for example in a doctoral dissertation written by Josef Schnippenkötter, a Jesuit physics teacher from Duisburg.[78] Reiche was among the few quantum physicists who referred to Nernst's theory, which he did by briefly dealing with the "radical" claim of a zero-point radiation filling all of space.[79] So did Richard Tolman in California, commenting on the ideas of Keesom, Nernst and Stern: "This 'nullpunkt energie' in the Nernst treatment is in equilibrium with radiant energy in the ether. On rise of temperature, energy is drawn not only from the surroundings but also from the reservoir of 'nullpunkt energie' and the principle of the conservation of energy becomes merely statistically true rather than true for the individual elements of the system."[80] It is, finally, worth mentioning that also Bohr was aware of Nernst's idea of 1916 that energy may not be conserved in an absolute but only statistical sense. In an unpublished manuscript from 1917 or 1918 he referred to "an interesting attempt to build up a theory on this basis [which] has been made by Nernst."[81] He was thus aware of Nernst's version of vacuum energy, but chose not to comment on it.

---

[75] Lodge (1920, p. 171). The linear dimension of Lodge's ether happened to be of the same order as the Planck length $(Gh/c^3)^{1/2} = 4 \times 10^{-33}$ cm or the length scale later appearing in string theory, which is about $10^{-32}$ cm. In Lodge (1907, p. 493), he illustrated the energy content of the ether more dramatically: "This is equivalent to saying that $3 \times 10^{17}$ kilowatt-hours, or the total output of a million-kilowatt power station for thirty million years, exists permanently, and at present inaccessibly, in every cubic millimetre of space."

[76] Valentiner (1919, p. 41).

[77] Günther (1924). Nernst (1928). On the instigation of Abram Joffe, a physicist at the University of St. Petersburg, in 1923 Nernst's *Weltgebäude* appeared in a Russian translation.

Apart from occasional references to Nernst's ideas of a vacuum zero-point radiation in the 1910s and 1920s, his hypothesis was effectively forgotten. Only much later, and in particular with the advent of dark energy, did it attract some attention. In a non-cosmological context the hypothesis reappeared in the late 1960s, when Timothy Boyer at the University of Maryland proposed a theory of electromagnetic zero-point energy that became one of the sources of the research program known as "stochastic electrodynamics." As Boyer pointed out in a paper of 1969, some features of his theory had been anticipated by Nernst more than fifty years earlier.[82]

---

[78] Schnippenkötter (1920, p. 18).
[79] Reiche (1918, p. 218) and Reiche (1921, p. 33).
[80] Tolman (1920, p. 1189), who saw no reason to adopt Nernst's theory or the idea of a zero-point energy.
[81] Bohr (1984, p. 15).

## 5  Some related contributions

In a couple of papers from 1925-1926 Stern studied from a thermodynamical point of view the conditions for radiation and elementary matter being in a state of cosmic equilibrium. At the time professor of experimental physics at the University of Hamburg, he was inspired by Arthur Eddington's recent theory of stellar evolution according to which the radiation energy from the stars was the result of matter-to-radiation nuclear processes, in the form of either proton-electron annihilation or fusion of hydrogen into helium.[83] In this context he referred to the possibility of inverse processes in which matter was produced by radiation energy: "In order to save the world from the heat death, Nernst once proposed the hypothesis that atoms of high atomic number might spontaneously be created by the radiation in cosmos [*Weltraumstrahlung*], to which he ascribed a zero-point energy."[84]

Stern considered a hollow space in equilibrium, meaning that the portion of matter radiated away in unit time would equal the amount of matter formed from the radiation. Although he did not explicitly introduce a cosmological perspective, he found it "very tempting to assume that cosmic space is in this state of equilibrium."[85] Stern's universe was a gigantic cavity filled with matter and radiation. Assuming the volume $V$ to be fixed and the matter particles of mass $m$ to behave like an ideal gas, he calculated the maximum entropy and in this way derived an expression for the number of particles $n$ per unit volume in equilibrium with blackbody radiation at temperature $T$:

$$ n = \frac{N}{V} = \frac{(2\pi m k T)^{3/2}}{h^3} exp\left( -\frac{mc^2}{kT} \right) $$

---

[82]  Boyer (1969). The zero-point energy of stochastic electrodynamics is not based on quantum mechanics, but has its origin in fluctuations of classical electromagnetic fields. Indeed, some advocates of stochastic electrodynamics see the research programme as a partial alternative to quantum mechanics, in the sense that quantum effects are attributed to the classical zero-point field.

[83]  Eddington (1920), and the authoritative monograph Eddington (1926, pp. 292-317). According to Eddington, the central temperature of a typical star would not exceed 30 million degrees.

[84]  Stern (1925, p. 448).

On account of the dominating effect of the exponential term, the concentration of particles comes out exceedingly small even at very high equilibrium temperatures. In addition, at a given temperature the number of protons will be much smaller than the number of electrons. According to Stern it would need a temperature of about 100 million degrees to support a particle density of one electron per $cm^3$, and for protons the temperature would be nearly 2000 times as great, $T \cong 10^{11}$ K. This obviously posed a problem, for not only were these temperatures much larger than what Eddington had calculated for the interior of stars (and thus implied an almost totally radiation-dominated universe), the result was also irreconcilable with the known electro-neutrality of matter: electrons and hydrogen nuclei had to be equally abundant, or very nearly so. One possible solution was in sight, but one that Stern chose to relegate to a footnote: "If any zero-point energy is to be ascribed to the radiation (Nernst) … [it] would lower the temperatures calculated."[86]

Stern's papers triggered some further work on the subject, in particular by Wilhelm Lenz and Pascual Jordan in Germany, by Richard Tolman and Fritz

---

[85]  Stern (1926a, p. 60), with English translation by H. Borns in Stern (1926b).
[86]  Stern (1926a, p. 62). See also Tolman (1934, pp. 147-151).

Zwicky in the United States, and by Seitaro Suzuki in Japan. Of these I shall pay particular attention to Lenz's little noticed contribution, which was the only one to refer to zero-point radiation. Lenz, a former student of Sommerfeld, had done important work in atomic and molecular theory and was at the time professor of theoretical physics at the University of Hamburg, thus a colleague of Stern. His paper of 1926 was directly inspired by Stern's works and also mentioned Nernst's hypothesis of a *Weltraumstrahlung*. Whereas Stern had not considered thermodynamics in relation to a particular cosmological model, Lenz applied similar reasoning to the favored relativistic model of the early 1920s, Einstein's closed and matter-filled universe proposed in 1917. (The other alternative, Willem de Sitter's model, was irrelevant since it contained no matter.) In Einstein's model, the radius of the universe $R$ was determined by the total mass $M$ according to

$$R = M \frac{\kappa}{4\pi^2}$$

Here $\kappa$ is Einstein's gravitational constant ($\kappa = 8\pi G/c^2$, where $G$ is Newtons's constant), which is related to the cosmological constant $\Lambda$ and the average density of matter $\rho$ by $\kappa\rho = 2\Lambda$. The volume of the universe is $V = 2\pi^2 R^3$. The first relation means that the radius of the universe grows with its mass, but as Lenz pointed out, "the radiation energy does not contribute to the expansion of the world," that is, to the increase of $R$.[87] He considered this to be an argument that weakened objections to a zero-point radiation in space:

> If one allows waves of the shortest observed wavelengths of $\lambda \cong 2 \times 10^{-11}$ cm (as in radioactive $\gamma$–rays) – and if this radiation, converted to material density ($u/c^2 \cong 10^6$), contributed to the curvature of the world – one would obtain a vacuum energy density of such as value that the world would not reach even to the moon.[88]

---

[87]  Lenz (1926, p. 643), who referred to information from the Austrian physicist Otto Halpern.

Lenz showed that if a particle of mass $m$ is created out of radiation, the radius and volume of the universe will increase by the quantities

$$\delta R = R\frac{m}{M} \quad \text{and} \quad \delta R = 3V\frac{m}{M}$$

Thus, the radius is changeable and only determined if there is a definite equilibrium between radiation and matter energy. This implied that the conditions underlying Stern's calculations had to be changed, and Lenz concluded that at equilibrium the radiation energy of the Einstein world must be equal to its matter energy. By means of the Stefan-Boltzmann radiation law he found that the temperature of the radiation would depend on the world radius as

$$T^2 = \frac{1}{R}\sqrt{\frac{2c^2}{\kappa a}},$$

where $a$ is the constant in the Stefan-Boltzmann law $\rho_{rad} = aT^4$. Expressing $R$ in cm, the expression can be written $T^2 \cong 10^{31}/R$. Lenz did not include a zero-point radiation energy in his calculations because of "the well-known uncertainties regarding this assumption." Arbitrarily assuming the radiation temperature to be 1 K, he was led to suggest a world radius of the order $10^{31}$ cm. Alternatively one might estimate the temperature from the radius, as given by the Einstein relation $R^2 = 2/\kappa\rho$. Taking from de Sitter the average density of matter in the universe to be $\rho \sim 10^{-26}$ g/cm$^3$, or $R \cong 10^{26}$ cm, Lenz arrived at the much too high space temperature 300 K. As to the question of electro-neutrality, that protons and electrons must be formed in equal numbers, he argued to have solved Stern's problem: "It makes no difference whether an electron or a hydrogen nucleus is formed, or whether they radiate away."

---

[88]  Ibid. The text has "β-rays," which must be a misprint and which I have consequently changed to "γ-rays."

The works by Stern and Lenz were reconsidered by Tolman at the California Institute of Technology, who in 1928 criticized some of Lenz's assumptions and derived formulae approximately agreeing with Stern's.[89] In a slightly later paper also Zwicky, at the time Tolman's colleague in Pasadena, took up the equilibrium approach pioneered by Stern. Zwicky concluded that Tolman's modification of Lenz's theory was "in clashing contradiction with the actual facts."[90] As mentioned, neither Tolman nor Zwicky considered the effect of a zero-point energy. In a paper of 1927 also Pascual Jordan, then at the University of Göttingen, developed the approach followed by Stern and Lenz. Applying the new forms of quantum statistics (Bose-Einstein and Fermi-Dirac) to the case where the total number of particles varies, he re-derived Stern's equilibrium formula. As a possible mechanism for matter-energy transformation in cosmic space Jordan mentioned proton-electron collisions of the kind

$$p^+ + 2e^- \rightarrow e^- + \gamma \, ,$$

which had recently been proposed by two American physicists.[91]

Stern presumably discussed the question of the gravitational effect of zero-point energy with Lenz in Hamburg, and we know that he discussed it with Pauli, who stayed in Hamburg between 1923 and 1928. As mentioned by Stern in his letter quoted in Section 3, for a period of time Pauli opposed the concept of zero-point energy, and he continued to deny the reality of such an energy in free space. According to the recollection of Pauli's two last assistents, Charles Enz and Armin Thellung, Pauli made an estimate of the gravitational effect of the zero-point radiation along the line of Nernst but with a cut-off of the classical electron radius $\lambda_{min} = e^2/mc^2 \cong 10^{-13}$ cm. He is said to have come to the conclusion that the radius of the Einstein universe would then "not even reach to the moon."[92] A recalculation made by Norbert Straumann, who followed some of the last lectures of Pauli, results in a world radius of 31 km, definitely confirming Pauli's estimate.[93] Interestingly, the conclusion reported by Lenz and Thellung is literally the same as given by Lenz in his 1926 paper, which may indicate that Pauli had discussed the issue with Lenz (which would have been natural) or at least that he was familiar with and many years later recalled Lenz's paper.

Within the context of the new Göttingen quantum mechanics, the first attempt to quantize the electromagnetic field was made by Pascual Jordan in the important *Dreimännerarbeit* from the fall of 1925, a work written jointly with Born and Heisenberg. Analyzing the field inside a cavity into a set of harmonic

Hamburg, but there is no indication that Lenz and Jordan discussed the problem of radiation in space during this period.

[92] First reported in Enz and Thellung (1960, p. 842), and later in Enz (2002, p. 152) and many other places, e.g., Rugh and Zinkernagel (2002). Pauli told the story to Enz and Thellung, and also to Stern about 1950, but it is unclear when he made the calculation.

[93] Straumann (2009) who says that he checked the calculation while a student in Zurich and after having heard about the problem from Enz and Thellung. Neither Enz, Thellung, Straumann nor other authors commenting on the story seem to be aware of Lenz's paper. This paper contains no mention of Pauli and there is also no indication of the problem in Pauli's scientific correspondence from the 1920s (Pauli 1979).

oscillators, Jordan assumed that in addition to what he called the "thermal energy" of the oscillators, there also had to exist a zero-point energy $\frac{1}{2}h\Sigma\nu_k$, where $k$ denotes the degrees of freedom.[94] In this way he was able to derive the fluctuation formula for blackbody radiation that Einstein had derived by statistical methods in 1909.

However, Jordan did not think of the field zero-point energy as physically real, for other reasons because of the infinite energy that would result from the infinitely many degrees of freedom of the field. "It is just a quantity of the calculation having no direct physical meaning," he wrote to Einstein at the end of 1925. "One can define physically only the thermal energy in the case of $T = 0$."[95] Einstein agreed, as he made clear in a letter to Ehrenfest a few months later:

> I have continued to concern myself very much with the Heisenberg-Born scheme. More and more I tend to the opinion that the idea, in spite of all the admiration for it, is wrong. A zero-point energy of cavity radiation should not exist. I believe that Heisenberg, Born and Jordan's argument in favour of it (fluctuations) is feeble.[96]

As to the infinity associated with the zero-point energy, Jordan soon found a way to get rid of it, namely by a substitution procedure which has been called "the first infinite subtraction, or renormalization, in quantum field theory."[97] The unphysical nature of the zero-point energy of space was spelled out in a paper he wrote jointly with Pauli and which appeared in early 1928:

> Contrary to the eigen-oscillations in a crystal lattice (where theoretical as well as empirical reasons speak to the presence of a zero-point energy), for the eigen-oscillations of the radiation no physical reality is associated to this "zero-point energy" of ½$h\nu$ per degree of freedom. We are here doing with strictly harmonic oscillators, and since this "zero-point energy" can neither be absorbed nor reflected – and that includes its energy or mass – it seems to escape any possibility for detection. For this reason it is probably simpler and more satisfying to assume that for electromagnetic fields this zero-point radiation does not exist at all.[98]

---

[94] Born et al. (1926), with English translation in van der Waerden (1967), where the relevant pages are pp. 377-385. It is well known that this part of the *Dreimännerarbeit* was due to Jordan (van der Waerden 1967, p. 55).

[95] Jordan to Einstein, 15 December 1925, quoted in Mehra and Rechenberg (1982-2000, vol. 6, p. 57).

[96] Einstein to Ehrenfest, 12 February 1926, quoted in Mehra and Rechenberg (1982-2000, vol. 4, p. 276).

[97] Schweber (1994, pp. 108-112).

In a review paper on the light quantum hypothesis from the same year, Jordan repeated that he did not believe in a vacuum zero-point energy. Characterizing the quantity as a "blemish" (*Schönheitsfehler*), he emphasized that it should be regarded "more as a formal complication than a real difficulty."[99]

A few years later, in an influential review of wave mechanics in the *Handbuch der Physik*, Pauli restated his and Jordan's belief that the zero-point energy could be ascribed to material systems only and not to the free electromagnetic field. It would, Pauli wrote, "give rise to an infinitely large energy per unit volume … [and] be unobservable in principle since it is neither emitted, absorbed nor scattered, hence it cannot be enclosed inside walls; and, as is evident from experience, it also does not produce a gravitational field."[100] In agreement with this view, in his *Handbuch* article Pauli wrote the expression for the energy density in such a way that the zero-point energy disappeared. Although Pauli did not refer to Nernst, and may not have read his lengthy paper of 1916, implicitly his arguments were a refutation of Nernst's dynamically active zero-point energy ether. However, Pauli disregarded the creation of matter particles out of vacuum or ether fluctuations, which was a crucial point in Nernst's hypothesis and also appeared in Lenz's work of 1926. If the vacuum field produces matter, it is no longer gravitationally inert.

---

[98]  Jordan and Pauli (1928, p. 154).
[99]  Jordan (1928, p. 195).
[100]  Pauli (1933, p. 250). See also Enz (1974) and Enz (2002, pp. 150-153, 181).

Apart from the proposals of Nernst and Lenz, the zero-point energy of quantum theory first appeared in a cosmological context in a note of 1930 by the American physicists Edward Condon and Julian Mack at the University of Minnesota. They phrased the problem in a manner Nernst might have approved of:

> According to quantum mechanics, a harmonic oscillator of frequency $\nu$ has a lowest energy state the energy of which is $\frac{1}{2}h\nu$. When the electromagnetic field is treated … as an assemblage of independent harmonic oscillators, one of which is associated with each of the normal modes of vibration of the ether, this leads to the result that there is present in all space an infinite positive energy density. It is infinite because there is supposed to be no upper limit to the frequencies of possible normal modes.[101]

The two physicists conjectured that the infinite energy density of space might just cancel the infinite negative energy density postulated by Paul Dirac in his new and controversial theory of the electron. According to Dirac, the infinite sea of negative energy states was itself unobservable but vacancies in it would appear as protons (or, in his later interpretation, as positive electrons).[102] Condon and Mack admitted that their "cosmological conjecture" was a speculation rather than a scientific theory. At any rate, neither they nor other physicists developed it further.

---

[101]  Condon and Mack (1930).

## 6  Steps toward dark energy

Classical ethers of the type assumed by Nernst and Wiechert were not the only ethers in the years about 1920. Surprisingly, on the face of it, Einstein began to speak of physical space, as described by the metrical tensor $g_{\mu\nu}$ in his general theory of relativity, as an "ether." In an address in Leiden in 1920 he stressed that "empty space" is not empty in the sense of having no physical properties. Quite the contrary, for space was indistinguishable from the gravitational field, which might be thought of as a non-absolute ether:

> According to the general theory of relativity, space is endowed with physical qualities; in this sense, therefore, there exists an ether. According to the general theory of relativity, space without ether is unthinkable; for in such space there not only would be no propagation of light, but also no possibility of existence for standards of space and time (measuring-rods and clocks), nor therefore any space-time intervals in the physical sense.[103]

Einstein was of the conviction that the concepts of space and ether had merged, and that the space-ether was primary relative to both matter and electromagnetic fields. Space, he said in a lecture at the University of Nottingham of 1930, "has in the last few decades swallowed ether and time and also seems about to swallow the field and the corpuscles, so that it remains the sole medium of reality."[104] Although quite differently justified, Einstein's ether had the feature in common with Nernst's version that it was physically active, indeed the source of all physical activity. However, Einstein never spoke of its activity as derived from a vacuum energy, as Nernst did, and he also did not relate its physical activity to the cosmological constant, as later physicists would do.

---

[102] For Dirac's "hole" theory of the electron, see Kragh (1990, pp. 88-105).
[103] Einstein (1920, p. 15), English translation in Einstein (1983, pp. 3-24). For a detailed account of Einstein's ethers, see Kostro (2000).

Suggestions of a connection between vacuum energy and the cosmological constant were absent until the 1930s, although a few physicists, among them Einstein and Hermann Weyl, considered the physical meaning of the constant. Einstein justified his introduction of the cosmological term $\Lambda g_{\mu\nu}$ not only as a means of keeping the universe in a static state, but also as a means of avoiding a cosmic negative pressure – for "experience teaches us that the energy density does not become negative."[105] Nearly thirty years later he elaborated: "The objection to this solution [the spatially finite, uniform world model] is that one has to introduce a negative pressure, for which there exists no physical justification. In order to make that solution possible I originally introduced a new member into the field equation instead of the above mentioned pressure."[106] It follows from the cosmological field equations, including a pressure term $p$, that

$$R^2 = (\Lambda - \kappa p)^{-1}$$

Without a cosmological constant this gives $R^2 = (-\kappa p)^{-1}$, which, in order to be positive, requires $p < 0$. If the matter pressure is zero and $\Lambda > 0$, as Einstein originally assumed, the result is instead $R^2 = 1/\Lambda$.

---

In a paper of 1919, in which he first expressed dissatisfaction with the cosmological constant – said to be "greatly detrimental to the formal beauty of the theory" – Einstein reconsidered the connection between the constant and a negative pressure.[107] However, the context was not cosmological, but an attempt at unification, namely, to provide a link between gravitation theory and the structure of electrical particles. Einstein considered an extended charged particle in the interior of which there was a negative pressure $(R_0 – R)$, where $R$ is the curvature scalar and $R_0$ its smaller value outside the particle. The negative pressure – a modernized version of the so-called Poincaré stress in classical electron theory[108] – was assumed to maintain the electromagnetic force in equilibrium. In regions where only electrical and gravitational forces were present, Einstein found that the cosmological constant could be expressed by the curvature as $4\Lambda = R$. He was at the time aware that the cosmological constant can formally be replaced by a negative pressure $p = - \Lambda/\kappa$, but without considering it important. Schrödinger had in 1918 suggested to change the expression for the energy-stress tensor $T_{\mu\nu}$ in a way that corresponded to an elimination of the cosmological term, and in a critical response Einstein pointed out that the proposal might be equivalent to introducing a negative pressure proportional to the cosmological constant.[109] However, neither Schrödinger nor Einstein entertained explicitly the idea of a vacuum energy with a corresponding negative pressure.

---

[107] Einstein (1919), with English translation in Einstein et al. (1952, pp. 189-198).

[108] Poincaré introduced in 1906 a non-electromagnetic, negative pressure acting only on the inside of the electron, where it balanced the repulsive electromagnetic forces tending to make the electron explode. The Poincaré stress was of the form $p = - e^2/8\pi R^4$, where $R$ is the radius of the electron. For a detailed examination of Poincaré's electron theory and the hypothesis of a negative pressure, see Miller (1973, especially pp. 297-301).

[109] Schrödinger (1918) and Einstein's response in the same volume on pp. 165-166. See also the editorial comment in Einstein (1998, p. 808, note 13).

That only came much later, although there were no particular reasons for the "delay." The interpretation of the cosmological constant as an effective vacuum energy density could have been made as early as 1917. Einstein wrote the cosmological field equations with the Λ-term as belonging to the space part of the equations, but if arranged as

$$R_{\mu\nu} - \frac{1}{2} g_{\mu\nu} R = -\kappa T_{\mu\nu} + \Lambda g_{\mu\nu} \, ,$$

they appear in a form with the cosmological term equivalent to a contribution to the energy-stress tensor $T_{\mu\nu}$. The contribution can be interpreted to imply a vacuum energy density $\rho_v$ associated with Λ and a corresponding negative pressure density proportional to the energy density. This appears even more clearly from the Friedmann equations, although these date from 1922 and became generally known only after 1930.[110] From these equations, written with both the cosmological constant and a pressure term, it follows directly that

$$\rho_v = \frac{\Lambda c^2}{8\pi G} \quad \text{and} \quad p_v = -\frac{\Lambda c^4}{8\pi G} \, ,$$

and then

$$p_v = -\rho_v c^2$$

In the parlance of later cosmologists, and with $c = 1$, the equation of state of the cosmological constant is given by the dimensionless parameter $w = p/\rho = -1$. (For ordinary matter, $w = 0$, and $w = 1/3$ for radiation). The energy density $\rho_v$ can be translated into a corresponding vacuum matter density as $\rho_v/c^2 = \Lambda/8\pi G$. In terms of the critical density introduced by Einstein and de Sitter in 1932 and given by

$$\rho_c = \frac{3H^2}{8\pi G} \, ,$$

where $H$ is the Hubble parameter, the vacuum energy density can be written as

$$\Omega_v = \frac{\rho_v}{\rho_c} = \frac{\Lambda}{3H^2}$$

---

[110] See, for example, Earman (2001, p. 192 and p. 206).

When the vacuum expands, the work done to expand it from volume $V$ to $V + dV$ is negative, namely, $pdV = -\rho c^2 dV$. In spite of the expansion, the energy density of the vacuum remains constant (while the energy increases). By the early 1930s it was well known that the effect of $\Lambda$ is equivalent to a negative pressure, which appears in some of the early reviews of the theory of the expanding universe.[111] The $\Lambda$-energy is sometimes described as a form of "anti-gravity," which is because the force of gravity, in the theory of general relativity, is determined by the combination $\rho + 3p/c^2$. The pressure term can usually be neglected, but in the case of the vacuum we have

$$\rho_v + \frac{3p_v}{c^2} = \rho_v - 3\rho_v = -2\rho_v \, ,$$

implying that gravity changes its sign.

---

[111] E.g., Zaycoff (1932), who concluded that in a contracting universe a final singularity could only be avoided if $\Lambda > 0$. The connection between the cosmological constant, the energy density and the negative pressure was also considered in Maneff (1932).

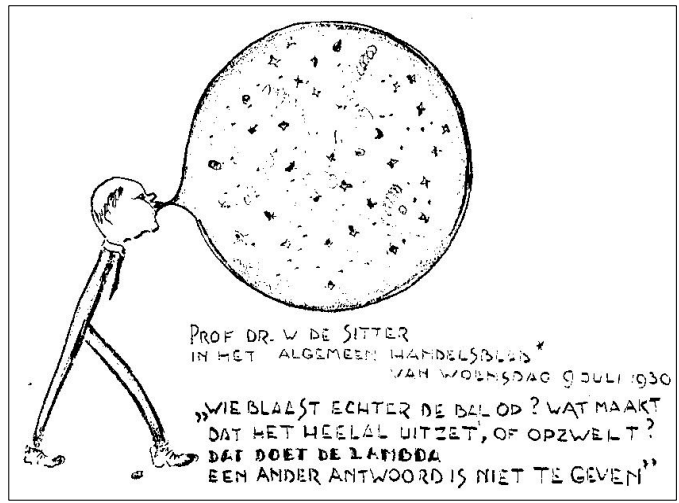

Figure 3. Sketch of Willem de Sitter (drawn as a "λ") in the *Algemeen Handelsblad* of 9 July 1930, as reproduced in Peebles (1993, p. 81). De Sitter says: "What, however, blows up the ball? What makes the universe expand or swell up? That is done by lambda. No other answer can be given."

The Dutch astronomer and cosmologist Willem de Sitter learned of the expanding universe in the early months of 1930. Although he knew that the expansion did not require a positive cosmological constant – there are expanding models with $\Lambda = 0$ – he believed that the constant was in fact responsible for the expansion of space. "What is it then that causes the expansion?" he asked in a popular article of 1931. His answer was that "the *lambda* does it":

It is the presence of *lambda*, the "cosmological constant" of Einstein, in the equations that not only closes up the universe, … but also provides the possibility of its changing its size. Why it expands and does not shrink, we do not know. … The expansion depends on the *lambda* alone. To some it may sound unsatisfactory that we are not able to point out the mechanism by which the *lambda* contrives to do it. But there it is, we cannot go beyond the mathematical equations, and … the behavior of *lambda* is not more strange or mysterious than that of the constant of gravitation *kappa*, to say nothing of the quantum-constant *h*, or the velocity of light *c*.[112]

The equations given above, relating the energy and pressure to the cosmological constant, were only explicitly stated in 1933, when the Belgian pioneer cosmologist (and father of the big bang) Georges Lemaître spent a period as guest professor at the Catholic University of America in Washington D.C. In a talk given to the U.S. National Academy of Sciences on 20 November 1933, he began by noting that, with a mean density of matter $\rho \cong 10^{-30}$ g/cm$^3$, "If all the atoms of the stars were equally distributed through space there would be about one atom per cubic yard, or the total energy would be that of an equilibrium radiation at the temperature of liquid hydrogen," that is, $T \cong 20$ K.[113] A few lines later he offered the following interpretation:

> Everything happens as though the energy in vacuo would be different from zero. In order that absolute motion, i.e., motion relative to vacuum, may not be detected, we must associate a pressure $p = -\rho c^2$ to the density of energy $\rho c^2$ of vacuum. This is essentially the meaning of the cosmical constant $\lambda$ which corresponds to a negative density of vacuum according to $\rho_0 = \lambda c^2/4\pi G \cong 10^{-27}$ g/cm$^3$.[114]

---

[112]  De Sitter (1931, pp. 9-10). See also Figure 3.

[113]  Lemaître (1934a, p. 12). Luminet (2007) calls this "a first intuition of a cosmic microwave background as a fossil radiation from the primeval atom," which is a misinterpretation. Lemaître did believe in a fossil radiation from the big bang, but he erroneously identified it with the cosmic rays.

[114]  Lemaître (1934a, p. 12). Notice that Lemaître's denominator was $4\pi G$ rather than $8\pi G$. Also, as noted by Earman (2001), he took $\Lambda > 0$ to correspond to a negative $\rho$, which may have been due to contemporary confusion about the sign conventions.

The negative density (and positive pressure) was not a slip of the pen, as we learn from a slightly later paper, where Lemaître said that the cosmological constant "may be regarded as equivalent to a density, of negative sign, and … accompanied with a positive pressure."[115] While Lemaître thus offered a physical interpretation of the cosmological constant as a vacuum energy density, he did not connect his interpretation with the zero-point energy of space or otherwise relate it to quantum physics. That a connection of this kind might exist seems to have been vaguely suspected by Weyl, who in a letter to Einstein of 1927 wrote: "All the properties that I had so far attributed to matter by means of $\Lambda$ are now to be taken over by quantum mechanics."[116] Alas, he did not elaborate.

Lemaître remained faithful to the cosmological constant as a vacuum energy throughout his life. In part inspired by Eddington, according to whom the "cosmical constant" was a manifestation of the quantum nature of the universe, he returned a few times to the subject, yet without attempting to clarify the quantum connection and without endorsing Eddington's unorthodox theory of cosmophysics.[117] For example, in an address given to the 1958 Solvay conference, the theme of which was astrophysics and cosmology, Lemaître stated that, "If some extension of relativity towards a broader field, such as quantum theory, has to be achieved the superfluous $\lambda$ term shall be very much welcomed." But instead of following up the idea, he merely remarked: "In the meantime, there is nothing to do than to use the cosmical term in astronomical applications."[118]

---

[115] Lemaître (1934b). See also Lemaître (1949), where he calls $\rho_0$ the "cosmological density" and stresses that the effect of the cosmological constant is to replace $\rho$ by ($\rho$ - $\rho_0$).

[116] Weyl to Einstein, 3 February 1927, quoted in Kerzberg (1989, p. 334).

[117] For Eddington's view of the cosmological constant as a measure of the zero from which energy and pressure are reckoned, see Eddington (1936, pp. 188, 195, and 258) and Earman (2001).

[118] Lemaître (1958, pp. 15-16). As pointed out in Rugh and Zinkernagel (2002), Lemaître was probably aware of the vacuum energy arising in quantum field theory at the time he gave his talk to the National Academy of Sciences.

Lemaître's insight attracted little attention and failed to inspire new work related to the strange form of vacuum energy and negative pressure. There seems to have been no mention of it through the 1930s and 1940s, possibly because cosmologists found Lemaître's interpretation unsurprising and because the cosmological constant was not considered the business of quantum theorists interested in the vacuum.[119]

The young Russian physicist Matvei Bronstein was among the few who (independently of Lemaître) entertained the idea of the cosmological constant as a form of cosmic energy. In a paper of 1933 he suggested that energy might be transferred between ordinary matter and the matter represented by the cosmological constant.[120] In modern parlance, the dark $\Lambda$–matter might decay. Bronstein was the first to suggest a time-varying constant $\Lambda = \Lambda(t)$, which he did "merely for the sake of generality." He considered the monotonic variation of $\Lambda$ to be an arrow of time in the sense that $\Lambda$ would always decrease and thereby explain why the universe is expanding rather than contracting. Moreover, the decreasing $\Lambda$ implied a violation of energy conservation, since the energy-conserving equation $dE + pdV = 0$ would not hold on a cosmic scale. Instead of energy conservation he obtained

$$\frac{dE}{dt} + p\frac{dV}{dt} = -\frac{\pi^2 c^2 R^3}{\kappa}\frac{d\Lambda}{dt}$$

---

[119]  According to the Web of Science database, Lemaître's 1934 paper has (until November 2011) been cited only 23 times, with 16 of the citations belonging to the period 1999-2011. The late attention to his work undoubtedly reflects the recent interest in dark energy. The Web of Science only mentions a single citing paper in the 1930s, and this paper, by George Gamow and Edward Teller, does not refer to the cosmological constant as a vacuum energy. It should be mentioned that the Web of Science is notoriously unreliable with respect to the older literature.

[120]  Bronstein (1933). See also Peebles and Ratra (2003, p. 571 and p. 577) and, for Bronstein's life and work, Gorelik and Frenkel (1994).

Bronstein justified his ideas by Bohr's contemporary ideas of energy nonconservation in nuclear and stellar physics, which made him suggest that the generation of radiant energy from the nuclei of stars did not satisfy the law of energy conservation. This radiant energy, he wrote at the end of his paper, is "formally equivalent to the introduction of *a new form of energy connected with the λ–field* which compensates Bohr's nonconservation."[121] Because Bronstein's Λ was evolving, he is sometimes mentioned as a precursor of "quintessence," a hypothetical form of dark energy that was introduced in 1998 as an alternative to the cosmological constant and which can vary in space and time. The equation of state for quintessence is $-1 < w \leq 0$.

It is outside the scope of the present paper to discuss in detail the later development, but a few comments relating to the 1950s and 1960s may be appropriate. Thus, Lemaître's negative cosmic pressure reappeared in the context of the version of steady-state theory proposed by William McCrea in the early 1950s. As an alternative to Fred Hoyle's "creation field," McCrea introduced a negative and non-observable negative pressure of the form $p = -\rho c^2$. This hypothetical pressure he described as a "zero-point stress" responsible for the creation of new matter in the expanding universe:

> According to relativity theory, the creation process must follow from the existence of a zero-point stress in space. Now the current quantum theory of fields endows space with several "virtual" zero-point properties. If any of these can be interpreted as producing a stress, it appears that the connexion might be established. (Such a treatment would require an examination of zero-point energy as well.)[122]

---

[121] Bronstein (1933, p. 82). Bronstein referred to Bohr (1932), where Bohr briefly suggested that energy conservation was violated in the interior of the stars. In fact, the suggestion was made several years earlier by Kramers, who in the semipopular book Kramers and Holst (1925, p. 140) argued that energy might be spontaneously generated in hot stars.

McCrea did not refer to either Lemaître's 1934 paper or the cosmological constant, although within his theory there was a formal similarity between the postulated cosmic stress and the cosmological constant. Nor did he, or other cosmologists at the time, mention the "Casimir effect" predicted on theoretical grounds in 1948 by the Dutch physicist Hendrik Casimir, a former assistant of Bohr and Pauli and at the time co-director of the research laboratory of the Philips Company. The Casimir effect is today generally understood as demonstrating the energy and negative pressure of empty space due to its zero-point energy, but for more than two decades the effect attracted little attention in quantum field theory and none in cosmology.[123] Inspired by McCrea's idea, the Polish physicist Jaroslav Pachner proposed in 1965 a cyclic model in the universe in which the singularity at the bounces ($R = 0$) was avoided by the building up of a negative vacuum pressure varying with the curvature of space. Since then, negative pressure has been a standard ingredient in cosmologies of the cyclic or bouncing type.[124]

The connection to the quantum mechanics of the vacuum that McCrea had vaguely anticipated was made explicit in works from 1965-1968 primarily by the Russian physicists Erast Gliner and Yakov Zel'dovich. Gliner, who worked at the Physico-Technical Institute in what was then Leningrad, seems to have been the first to suggest that the universe might have been begun its expansion in a vacuum-like state, an idea which eventually was developed into the inflationary scenario.[125] He called the hypothetical form of vacuum-like matter a "μ-vacuum" and ascribed to it a negative pressure. Gliner argued that the hypothesis of a negative-pressure vacuum was not "utterly unrealistic" because attempts to describe the structure of elementary particles "would lead to the conclusion that inside the particle there must be a negative pressure which balances the electrostatic repulsion."[126] Probably without knowing it, his remark connected to Einstein's theory of 1919 and the even earlier idea of Poincaré concerning the structure of the electron.

---

[122]  McCrea (1951, pp. 573-574). On McCrea's theory and its relationship to the ordinary theory of general relativity, see Kragh (1999).

[123]  For a critical discussion of the Casimir effect, see Rugh et al. (1999). McCrea (1986) discussed the Casimir effect and its relation to the cosmological "substratum radiation" in his Milne Lecture of 1985.

[124]  Pachner (1965). Kragh (2009).

Interpreting the vacuum energy of empty space as the result of quantum fluctuations in the zero-energy field, in papers of 1967 and 1968 Zel'dovich, at the Institute of Applied Mathematics in Moscow, pointed out that "we can speak of an energy density of the vacuum and a pressure (stress tensor) of the vacuum."[127] Moreover, by assuming a cut-off corresponding to the mass of a proton he derived a zero-point energy $\rho_v$ of the order $10^{17}$ g/cm$^3$, noting that it much exceeded the observational bound on the cosmological constant. This was the beginning of the "cosmological constant problem," namely, that the cosmological constant as calculated from the zero-point energy density of the vacuum $\Lambda_{QFT}$ is hugely larger than bounds imposed by observation. Zel'dovich compared the calculated $\Lambda_{QFT} \cong 10^{-10}$ cm$^{-2}$ with the observational limit $\Lambda_{obs} < 10^{-54}$ cm$^{-2}$. Although considering the first value to have "nothing in common with reality," he nonetheless suggested that the cosmological constant might arise from the vacuum of quantum field theory. In fact, the discrepancy between the two versions of the cosmological constant soon turned out to be much larger than estimated by Zel'dovich.

---

[125] The work of Gliner, Zel'dovich and other Russian theorists anticipated the early theories that led to the inflation scenario (due to Alexei Starobinsky and Allan Guth) and turned the vacuum energy into a concept of particular relevance for the very early universe. For this connection, see Smeenk (2005).

[126] Gliner (1966, p. 381), published in Russian 1965. Gliner seems to have been unaware of the electron model suggested by Casimir in 1953 and in which the Poincaré stress was explained as an effect of the zero-point energy of the electromagnetic field. On this model, see Carazza and Guidetti (1986).

[127] Zel'dovich (1968, p. 382). This paper has been republished, with an editorial introduction by Varun Sahni and Andrzej Krasinski, in *General Relativity and Gravitation* 40 (2008): 1557-1591.

## 7 Conclusion

Apart from reviewing the early theory of zero-point energy and its historical links to half-integral quantum numbers, in this paper I have drawn attention to the first attempts of applying ideas of quantum theory to cosmic space. In this respect Nernst's theory of 1916 should be counted as a pioneering work, and that in spite of its speculative nature and basic assumption of an ether filling up the universe. However, Nernst's ideas were out of tune with mainstream physics and for this reason alone exerted little influence on the further development. In later works, at least indirectly inspired by Nernst, the Hamburg physicists Stern and Lenz investigated the equilibrium of matter and radiation in the universe, which made Lenz to estimate the temperature of space according to Einstein's model of a closed universe. Lenz's paper of 1926 is little known and has been ignored by most historians of physics and cosmology. It deserves better.

With the exception of Nernst, the zero-point energy of free space was an unwelcome concept that found no place in quantum physics until the 1930s. Moreover, it remained isolated from the vacuum energy associated with the cosmological constant also after 1934, when Lemaître clearly formulated the connection between vacuum energy, negative vacuum pressure, and the cosmological constant. This insight, which did not rely specifically on the expanding universe, could have been stated many years earlier. But it was not, and when it was stated it attracted no interest. Although the cosmological constant is mathematically equivalent to the gravitational effects of vacuum energy, conceptually the two quantities are entirely different: while the first is a property of space, the latter is a quantum effect. It was only after the establishment of modern big bang theory in the mid-1960s that Zel'dovich thought of integrating the quantum-mechanical zero-point energy with the vacuum energy of the cosmological constant, thereby starting a line of development that would lead to the famous cosmological constant problem and give the vacuum energy a central role in cosmological research.

Vacuum energy in the form of the cosmological constant appeared as a crucial element in the inflation scenarios of the early 1980s, but limited to the very early universe. Lemaître's version of vacuum energy, on the other hand, had an effect that became relatively more important as the expansion proceeds, and in this sense it was closer to the dark energy of modern cosmology. Although dark energy came as an observational surprise in the late 1990s, it was not a complete surprise. A few cosmologists, and perhaps first Hans-Joachim Blome and Wolfgang Priester from Germany, realized in the 1980s that the universe might be in a state dominated by vacuum energy and have been so for several billion years (Figure 4).[128]

**Acknowledgments**   I would like to thank Dominique Lambert for useful comments on an earlier version of this paper.

---

[128] Blome and Priester (1984), who not only referred to the works of Gliner and Zel'dovich, but also to McCrea (1951) and Einstein (1920). On the other hand, they seem to have been unaware of Nernst's early anticipation of dark energy. Blome and Priester (1985) concluded in favour of a vacuum energy of $\rho_v c^2 = \Lambda c^4 / 8\pi G \approx 10^{-8}$ erg/cm³.  On Priester's works in cosmology, including his fascination of the cosmological constant, see Overduin et al. (2007).

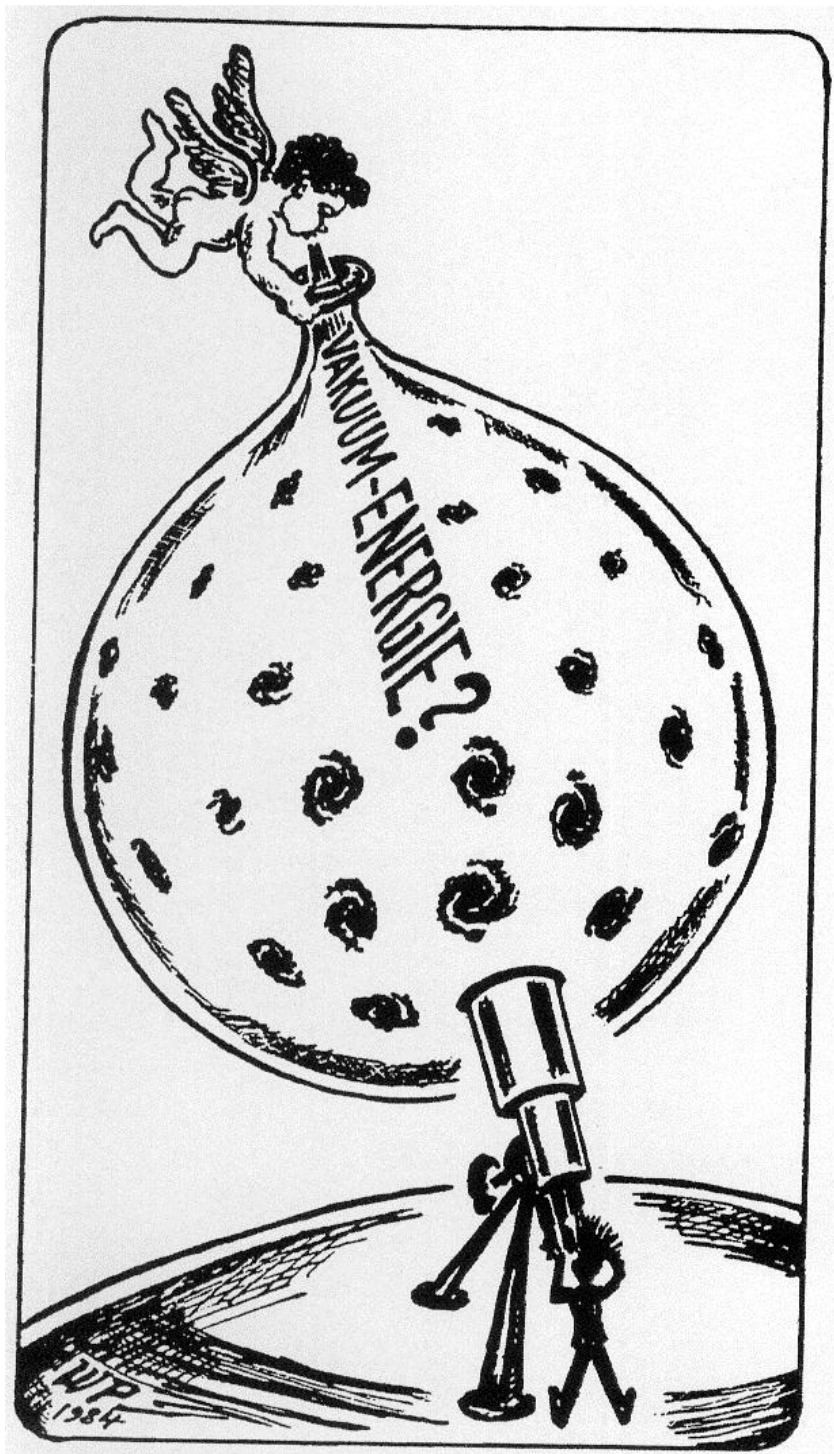

Figure 4. Wolfgang Priester's illustration of vacuum energy from 1984, as reproduced in Overduin et al. (2007, p. 419). Source: Priester (1984).